\documentclass[fleqn,usenatbib]{mnras}

\usepackage{newtxtext,newtxmath}

\usepackage[T1]{fontenc}

\DeclareRobustCommand{\VAN}[3]{#2}
\let\VANthebibliography\thebibliography
\def\thebibliography{\DeclareRobustCommand{\VAN}[3]{##3}\VANthebibliography}

\usepackage{graphicx}	
\usepackage{amsmath}	
\usepackage[dvipsnames]{xcolor}
\usepackage{bigints}
\usepackage{subfigure}
\usepackage[normalem]{ulem}
\usepackage{float}
\usepackage{hyperref}
\usepackage{natbib}

\defcitealias{Leonard2018}{L2018}
\defcitealias{Fortuna2021}{F2021}
\defcitealias{LSST_SRD}{LSST DESC 2018}

\title[IA from multiple shear estimates]{Intrinsic alignment from multiple shear estimates: A first application to data and forecasts for Stage IV}

\author[C. MacMahon-Gell\'er et al.]{
Charlie MacMahon-Gell\'er$^{1}$\thanks{E-mail: \url{c.macmahon@ncl.ac.uk}}
and C. Danielle Leonard$^{1}$
\\
$^{1}$School of Mathematics, Statistics and Physics, Herschel Building, Newcastle University, Newcastle-upon-Tyne, NE1 7RU, United Kingdom
}

\date{Accepted 2023 December 15. Received 2023 December 6; in original form 2023 June 20}

\pubyear{2023}

\begin{document}
\label{firstpage}
\pagerange{\pageref{firstpage}--\pageref{lastpage}}
\maketitle

\begin{abstract}
Without mitigation, the intrinsic alignment (IA) of galaxies poses a significant threat to achieving unbiased cosmological parameter constraints from precision weak lensing surveys. Here, we apply for the first time to data a method to extract the scale dependence of the IA contribution to galaxy-galaxy lensing, which takes advantage of the difference in alignment signal as measured by shear estimators with different sensitivities to galactic radii. Using data from Year 1 of the Dark Energy Survey, with shear estimators \texttt{METACALIBRATION} and \texttt{IM3SHAPE}, we investigate and address method systematics including non-trivial selection functions, differences in weighting between estimators, and multiplicative bias. We obtain a null detection of IA, which appears qualitatively consistent with existing work. We then forecast the application of this method to Rubin Observatory Legacy Survey of Space and Time (LSST) data and place requirements on a pair of shear estimators for detecting IA and constraining its 1-halo scale dependence. We find that for LSST Year 1, shear estimators should have at least a $40\%$ difference in IA amplitude, and the Pearson correlation coefficient of their shape noise should be at least $\rho=0.50$, to ensure a $1\sigma$ detection of IA and a constraint on its 1-halo scale dependence with a signal-to-noise ratio greater than $1$. For Year 10, a $1\sigma$ detection and constraint become possible for $20\%$ differences in alignment amplitude and $\rho=0.50$.
\end{abstract}

\begin{keywords}
gravitational lensing: weak - galaxies: interactions - galaxies: haloes - galaxies: statistics - large scale structure of Universe - cosmology: theory.
\end{keywords}



\section{Introduction}
\label{sec:intro}

Einstein's theory of General Relativity predicts that massive objects alter the geometry of surrounding space-time and consequently affect the path of light-rays passing close to them, in an effect known as gravitational lensing. On large scales, the subtle lensing of light-rays from background source galaxies by foreground lens masses is only detectable by correlating the shapes of many galaxies, as their intrinsic ellipticity is much larger than any observed change (shear) induced by large-scale lensing. This effect, referred to as \textit{weak gravitational lensing}, has proven to be an important scientific tool for probing the matter distribution of the Universe, which in turn has allowed us to develop our understanding of dark matter and dark energy (e.g.~\citealp{Hu2002,Baldauf2010,Weinberg2013,Abbott2018}).

Weak lensing measurements typically consider auto-correlations between the shapes of source galaxies (\textit{cosmic shear}; CS), or cross-correlate the shapes of source galaxies with the positions of foreground lens galaxies (\textit{galaxy-galaxy lensing}; GGL) (for a review on the theory of weak lensing, see~\cite{Bartelmann_2001}). However, both cosmic shear and galaxy-galaxy lensing are susceptible to contamination by correlations known as \textit{intrinsic alignments} (IA), resulting from tidal physics as well as galaxy formation and evolution effects. The origin of IA at different scales and for different galaxy types is an active area of research, but is generally understood to include tidal effects, as well as possible contributions from galaxy evolution history and environment \citep{Croft2000,Heavens2000,Troxel_2015}. In GGL, which will be the focus of this work, correlations due to IA exist between lens and source galaxies before the weak lensing effect imprints on our observations. Consequently, if IA is not properly accounted for, it can result in biased lensing estimates and thus biased constraints on cosmological models.

Current Stage III surveys - such as the Dark Energy Survey (DES;~\citealp{DES}), Kilo-Degree Survey (KiDS;~\citealp{KiDS}), and Hyper Suprime-Cam survey (HSC;~\citealp{HSC}) - and upcoming Stage IV surveys - such as the Rubin Observatory's Legacy Survey of Space and Time (LSST;~\citealp{Ivezic2019}), and Euclid~\citep{Euclid2022} - are vastly decreasing the statistical uncertainty on weak lensing measurements. With such a wealth of modern surveys providing greater statistical power, IA is becoming a significant source of uncertainty (e.g.~\citealp{Samuroff2019, Secco2022}) and is forecast to become even more so in the near future~\citep{Krause2016}. 

Direct detection of intrinsic alignment, using smaller samples of more luminous `source' galaxies for which spectroscopic redshifts are available, has provided key insight into the physics of intrinsic alignment, by directly selecting only those `source' and `lens' galaxies which are truly physically associated \citep{Mandelbaum2006, Hirata_2007, Okumura2009, Singh2015}. However, in order to achieve the statistical power needed to make an accurate weak lensing measurement, millions of galaxy images are required. Thus, many surveys measure galaxy redshifts using \textit{photometry} with much broader spectral bands than spectroscopy, leading to larger associated uncertainties. Such surveys also do not always have a representative spectroscopic sub-sample available. 

Other methods for measuring or mitigating the IA contamination to GGL have exploited the redshift dependence of the effect using methods such as binning sources in photometric redshift (photo-z), to separate those which are closer or further in redshift from the lenses~\citep{Heymans2004,Hirata2004b,Joachimi2011,Blazek2012}. However, such methods could be impacted by potentially large photo-z uncertainties, which in the worst cases may even be incorrectly estimated~(e.g.~\cite{Bernstein2010}).

Advances are being made in the measurement of photo-z (e.g.~\citealp{Bilicki2018}), and it may soon be possible to obtain a large enough spectroscopic sample to carry out high-precision weak lensing studies (e.g.~\citealp{DESI}). Nonetheless, characterising the associated uncertainties remains an important consideration for upcoming photometric lensing surveys, such as LSST and Euclid. Novel methods to measure the IA contamination are also being proposed to address the issues associated with photo-z, such as the self-calibration methods (targeting cosmic-shear;~\citealp{Zhang2010,Troxel2012,Yao2017,Yao2019}). However, these methods are still somewhat dependent on the photo-z uncertainty, although do include parameters to account for this. 

In \cite{Leonard2018} (hereafter: \citetalias{Leonard2018}), a novel method for measuring and / or constraining the scale dependence of intrinsic alignment was proposed. This method attempts to use the dependence of the IA signal's amplitude on the radial scales within a galaxy, rather than its dependence on redshift. We expect the outer radial regions of galaxies to be more aligned with local structure than the inner radial regions, which results in twisting of the isophotes in a galaxy's light profile. Using observational data,~\cite{Singh2016} showed that shear estimators with sensitivity to different radial scales within galaxies, had different levels of IA contamination, due to isophotal twisting theorised to result from IA. Further observational evidence was shown in~\cite{Georgiou2019a}, where altering the radial weighting of a shape estimator resulted in different measured alignment amplitudes. \cite{Tenneti2014} also showed this effect in simulated data. 

The method of \citetalias{Leonard2018} (henceforth referred to as the multi-estimator method; MEM) therefore looks to compare weak lensing measurements from two different shear estimators, with sensitivity to different radial regions of galaxies. If the lensing contribution to shear can be shown to be the same in both estimators (as should ideally be the case), taking the difference of the two estimates would `cancel out' the lensing signal (since the lensing effect does not depend on the radial region of the galaxy from which the light originated). This leaves a portion of the IA signal, determined by the difference in IA amplitude between the radial regions of the galaxy which each estimator probes. The cancellation of the lensing signal requires some assumptions, which will be discussed in greater detail later as a key subject of this work.

The advantages of such a method are twofold. Firstly, since the lensing signal is cancelled, it does not need to be measured and removed. This may make such a method particularly robust in the case of catastrophic photo-z error estimations, which could become more likely as upcoming surveys image further and fainter sources than ever before. Secondly, correlations in shape noise and cosmic variance between the two estimators could reduce the uncertainty in the measured IA signal. This would allows us to test IA at small scales within the 1-halo regime, which current models struggle to describe due to non-linear effects. 

In this work, we focus on GGL and do not consider the MEM in the context of direct application to cosmic shear, as non-local gravitational-intrinsic correlation would significantly complicate the formalism of the estimator. Therefore, we consider the simplest scenario for the estimator (GGL) while the method is still in development, with the view of potentially extending to direct cosmic shear applications in future work.

This paper is structured as follows: in Section \ref{sec:theory}, we review the formalism of the MEM from \citetalias{Leonard2018} and introduce a series of assumptions made in its most basic construction. We then go on to derive a new fundamental expression in the absence of one of these assumptions and discuss other potential complications. In Section \ref{sec:case_study}, we present the first observational measurement with MEM, using the Dark Energy Survey Year 1 galaxy shape catalogues, and propose methods to address the complications to the basic formalism outlined in Section \ref{sec:theory}. In Section \ref{sec:forecasting}, we consider the method in the context of upcoming Stage IV lensing surveys and carry out forecasts to place requirements on shear estimators for use with the MEM. Finally, in Section \ref{sec:conclusion}, we conclude by placing these results in the context of classes of shear estimators planned for deployment in Stage IV surveys.

\section{Theory}
\label{sec:theory}

\subsection{Basic method formalism}
\label{sec:formalism}

In this section, we briefly review the mathematical formalism of the MEM, as introduced in \citetalias{Leonard2018}.

The key GGL observable considered here is tangential shear, which measures the level of alignment and ellipticity distortion tangential to a lens. We consider two different estimators for the tangential shear, as given by two different shear estimation methods, $\gamma_{\rm t}$ and $\gamma_{\rm t}^\prime$,
\begin{align}
    \Tilde{\gamma}_{\rm t}(\theta) &=  B(\theta)(1+m)\left( \frac{\overset{\rm lens}{\underset{j}{\sum}} \Tilde{w}_{j} \, \Tilde{\gamma}_j} {\overset{\rm lens}{\underset{j}{\sum}}\Tilde{w}_j} \right) \nonumber \\ &= B(\theta)(1+m) \left( \frac{\overset{\rm lens}{\underset{j}{\sum}} \Tilde{w}_{j} \, \gamma_{\rm L}^j} {\overset{\rm lens}{\underset{j}{\sum}}\Tilde{w}_j} + \frac{\overset{\rm lens}{\underset{j}{\sum}} \Tilde{w}_{j} \, \gamma^j_{\rm IA}} {\overset{\rm lens}{\underset{j}{\sum}}\Tilde{w}_j} \right),
    \label{eqn:1}
\end{align}

\begin{align}
    \Tilde{\gamma}^{\prime}_{\rm t}(\theta) &=  B(\theta)(1+m^\prime)\left( \frac{\overset{\rm lens}{\underset{j}{\sum}} \Tilde{w}_{j} \, \Tilde{\gamma}_j^\prime} {\overset{\rm lens}{\underset{j}{\sum}}\Tilde{w}_j} \right) \nonumber \\ &= B(\theta)(1+m^\prime) \left( \frac{\overset{\rm lens}{\underset{j}{\sum}} \Tilde{w}_{j} \, \gamma_{\rm L}^j} {\overset{\rm lens}{\underset{j}{\sum}}\Tilde{w}_j} + \frac{a \overset{\rm lens}{\underset{j}{\sum}} \Tilde{w}_{j} \, \gamma^j_{\rm IA}} {\overset{\rm lens}{\underset{j}{\sum}}\Tilde{w}_j} \right).
    \label{eqn:2}
\end{align}
Here, the label `lens' indicates a sum over lens-source pairs. $\gamma$ denotes shear, subscripts L and IA denote lensing and IA contributions respectively, and a tilde denotes an observed quantity. $\theta$ is the on-sky lens-source angular separation, $\Tilde{w}_j$ are weights given to each lens-source pair, and $a$ is a factor which scales the IA amplitude of one estimator relative to the other. $m$ and $m^\prime$ represent sample-level multiplicative bias, residual in the estimators post-calibration (note these are not the full multiplicative bias values, only a portion of the bias that remains due to uncertainty on the calibration values). 

The boost factor, $B(\theta)$, accounts for `excess' galaxies, which are physically associated with the lens due to clustering, and as such not expected to be physically associated in a random sample. It is given by:
\begin{equation}
    B(\theta) = \frac{N_{\rm rand}}{N_{\rm lens}} \times \frac{\overset{\rm lens}{\underset{j}{\sum}}\Tilde{w}_j}{\overset{\rm rand}{\underset{j}{\sum}}\Tilde{w}_j},
    \label{eqn:3}
\end{equation}
where `rand' indicates a sum over random-source pairs, i.e. sources paired with galaxies from a catalogue of random points drawn from the lens redshift distribution. $N_{\rm rand}$ is the number of randoms and $N_{\rm lens}$ is the number of lenses.

We now apply the assumptions that the weights for the two methods are identical, and the multiplicative biases residual after calibration, $m$ and $m^\prime$, are demonstrably subdominant. We will later revisit these assumptions in detail. Taking the difference of our estimators, by subtracting equation \ref{eqn:2} from equation \ref{eqn:1}, now gives:
\begin{equation}
    \Tilde{\gamma}_{\rm t}(\theta) - \Tilde{\gamma}_{\rm t}^\prime(\theta) = (1-a)\frac{\overset{\rm lens}{\underset{j}{\sum}} \Tilde{w}_j\gamma^j_{\rm IA}}{\overset{\rm rand}{\underset{j}{\sum}}\Tilde{w}_j}.
    \label{eqn:4}
\end{equation}
Now consider a sample of lens-source pairs in which the lens and source have small enough line-of-sight separation that we would expect them to be intrinsically aligned. The scale in which it is conventionally assumed IA could be present is $100$ Mpc/h~(e.g. \citetalias{Leonard2018}) and we adopt this assumption here. The quantity of interest to us is the tangential shear due to IA per contributing lens-source pair. To account only for contributing pairs, we divide equation \ref{eqn:4} by the sum of the weights of contributing pairs. To express this, we make the following definition:
\begin{equation}
    \frac{\overset{\rm rand}{\underset{j}{\sum}}\Tilde{w}_j}{\overset{\rm excess}{\underset{j}{\sum}}\Tilde{w}_j + \overset{\rm rand,close}{\underset{j}{\sum}}\Tilde{w}_j} = \frac{1}{B(\theta) - 1 + F},
    \label{eqn:5}
\end{equation}
with F defined by:
\begin{equation}
    F \equiv \frac{\overset{\rm rand,close}{\underset{j}{\sum}}\Tilde{w}_j}{\overset{\rm rand}{\underset{j}{\sum}}\Tilde{w}_j},
    \label{eqn:6}
\end{equation}
where `\textrm{rand, close}' denotes a sum over sources within $\Pi = 100$ Mpc/h line-of-sight separation of random points~\citepalias{Leonard2018}. The choice of $100$Mpc/h does warrant further investigation, which we touch on again in Section \ref{sec:forecasting} below.

Finally, multiplying equation \ref{eqn:4} by equation \ref{eqn:5} gives us an expression to extract a portion of the IA signal, per intrinsically aligned lens-source pair, $(1-a)\Bar{\gamma}_{\rm IA}$:
\begin{equation}
    (1-a)\Bar{\gamma}_{\rm IA}(\theta) = \frac{\Tilde{\gamma}_{\rm t}(\theta) - \Tilde{\gamma}^\prime_{\rm t}(\theta)}{B(\theta) - 1 + F}.
    \label{eqn:7}
\end{equation}
Equation \ref{eqn:7} is the fundamental equation of this method. From it we can measure a portion of the IA contribution up to an amplitude determined by $a$. For example, a value of $a=0.8$ indicates a $20\%$ difference the IA contamination of our two estimators. Even in the case where we only recover a small fraction of the IA contamination, provided the signal is above zero, this gives us the ability to extract information about the scale dependence of intrinsic alignment, potentially inside the non-linear 1-halo regime. If an estimate of $a$ could be obtained as the radial sensitivities of the estimators were known, then it would be possible to fully model the IA contribution to tangential shear for either of these estimators. However, even if $a$ could not be estimated, information on the scale dependence would allow for the amplitude of an IA model to be calibrated within cosmological parameter estimation pipelines. A more detailed discussion on the significance of $a$ is given in Section \ref{sec:a}.

While equation \ref{eqn:7} represents the ideal case of this method, as mentioned above, we have made several strong assumptions about the relative characteristics of the shear estimation methods to obtain it. We now go beyond the method as introduced in \citetalias{Leonard2018}, to explore the consequences of relaxing these assumptions.

\subsection{The effect of residual multiplicative bias}
\label{sec:mbias_formalism}
The work of \citetalias{Leonard2018} assumed residual multiplicative bias due to uncertainty in the multiplicative bias calibration to be subdominant and thus ignored. This is because \citetalias{Leonard2018} was considering future shear estimation methods with demonstrably subdominant calibration uncertainty, such as the Bayesian Fourier Domain method proposed by~\cite{Bernstein2014}. In the case where the uncertainty on multiplicative bias calibration cannot be shown to be subdominant, residual multiplicative bias can remain in the estimators, and equation \ref{eqn:7} must be re-formulated to include terms accounting for the residual bias in each estimator. Subtracting equation \ref{eqn:2} from equation \ref{eqn:1} in this instance yields:
\begin{multline}
    \tilde{\gamma}_{\rm t} - \tilde{\gamma}_{\rm t}^\prime
    = B(\theta)
    \left[
    (m - m^\prime)
    \frac{\overset{\rm lens}{\underset{j}{\sum}}\tilde{w}_{j}\gamma^{j}_{\rm L}}{\overset{\rm lens}{\underset{j}{\sum}}\tilde{w}_{j}} \right. \\ \left.
    +
    (m - a m^\prime)
    \frac{\overset{\rm lens}{\underset{j}{\sum}}\tilde{w}_{j}\gamma^{j}_{\rm IA}}{\overset{\rm lens}{\underset{j}{\sum}}\tilde{w}_{j}}
    +
    (1-a)
    \frac{\overset{\rm lens}{\underset{j}{\sum}}\tilde{w}_{j}\gamma^{j}_{\rm IA}}{\overset{\rm lens}{\underset{j}{\sum}}\tilde{w}_{j}}
    \right].
    \label{eqn:8}
\end{multline}
Normalising by the weighted number of physically associated pairs gives,
\begin{multline}
    \frac{\tilde{\gamma}_{\rm t} - \tilde{\gamma}_{\rm t}^\prime}{B(\theta) - 1 + F} = 
    (m - m^\prime) \Bar{\gamma}_{\rm L,PA}
    +
    (m - am^\prime) \Bar{\gamma}_{\rm IA}
    +
    (1-a) \Bar{\gamma}_{\rm IA},
    \label{eqn:9}
\end{multline}
where $\Bar{\gamma}_{\rm L,PA}$ and $\Bar{\gamma}_{\rm IA}$ respectively represent the average lensing and IA contributions to tangential shear, per lens-source pair. Here, the subscript ${\rm PA}$ denotes the residual is normalised by the weighted number of physically associated pairs, not that only those pairs have contributed to this lensing signal.

Essentially, this implies the lensing contribution to shear does not fully cancel, in the case where residual multiplicative bias in the estimators is not subdominant, leaving a lensing residual, $(m - m^\prime) \Bar{\gamma}_{\rm L,PA}$.  Note that the term `lensing residual' refers generally to any part of the lensing signal that was not cancelled by taking the difference of the two tangential shears, whereas residual multiplicative bias specifically refers to a bias remaining in the tangential shear estimates due to the uncertainty on the multiplicative bias calibration.

Due to the percent level contribution of the IA signal to the full tangential shear, even percent level uncertainty on bias calibration has the potential to leave a lensing residual which dominates the IA signal in the MEM. Accounting for multiplicative bias uncertainty in the case where it is not subdominant is therefore imperative to the success of this method.

\subsection{Weighted source redshift distributions}
\label{sec:weights_formalsim}
Another potential source of a lensing residual arises when our two estimators have different weighting schemes. It is clear from equations \ref{eqn:1} and \ref{eqn:2} that different weights would result in different values of tangential shear, even in the absence of IA and multiplicative bias uncertainty, as well as different boost (equation \ref{eqn:3}) and $F$ (equation \ref{eqn:6}) values.

To see how the different weighting schemes can manifest as different tangential shears, it is simplest to express $\Tilde{\gamma}_{\rm t}$ as a Fourier space integral over the matter power spectrum. Following, for example, ~\cite{Prat_2018}, the tangential shear is given by,
\begin{align}
    \nonumber \Tilde{\gamma}_{\rm t}(\theta)&= b \frac{3}{2} \Omega_m \left( \frac{H_0}{c} \right)^2
    \int\frac{d\ell}{2\pi} \ell J_2(\theta\ell) \\ 
    &\times \int dz \Bigg[\frac{g(z)n_{\rm L}(z)}{a(z)\chi(z)} P_{\delta\delta}\left(k = \frac{\ell}{\chi(z)}, \chi(z)\right)\Bigg],
    \label{eqn:10}
\end{align}
Here, we have assumed lens galaxies trace the underlying matter with a linear bias, such that $\delta_{\rm g} = b \, \delta_{\rm m}$. $J_2$ is the second order Bessel function, $\ell$ is the angular wavenumber, $k$ is the 3D wavenumber, $a$ is the scale factor (not the IA offset parameter defined previously), $\chi$ is comoving distance, and $n_{\rm L}(z)$ is the lens redshift distribution. The quantity of interest that changes depending on the weighted redshift distribution of the sources is $g(z)$, the lensing efficiency, given by:
\begin{align}
    g(z) = \int^{\infty}_z dz^\prime \Tilde{w}(z) n_{\rm s}(z) \frac{\chi(z^\prime)-\chi(z)}{\chi(z^\prime)},
    \label{eqn:lens_effic}
\end{align}
where we have given the weights as a function of redshift as, typically, fainter and noisier galaxies are more likely to be observed at higher redshifts and have lower associated weights. From equation \ref{eqn:lens_effic} it is apparent that a redshift distribution weighted by two different schemes for each estimator (i.e. $\Tilde{w}(z)n_{\rm s}(z)$), will result in different lensing efficiencies and thus different tangential shears. This is therefore another potential source of a lensing residual, which could contaminate attempts to measure the IA amplitude offset between the two estimators, if not adequately corrected for.

\subsection{Galaxy size and limiting resolution}
\label{sec:radial_weight_res}
\begin{figure}
    \centering
    \includegraphics[width=0.5\textwidth]{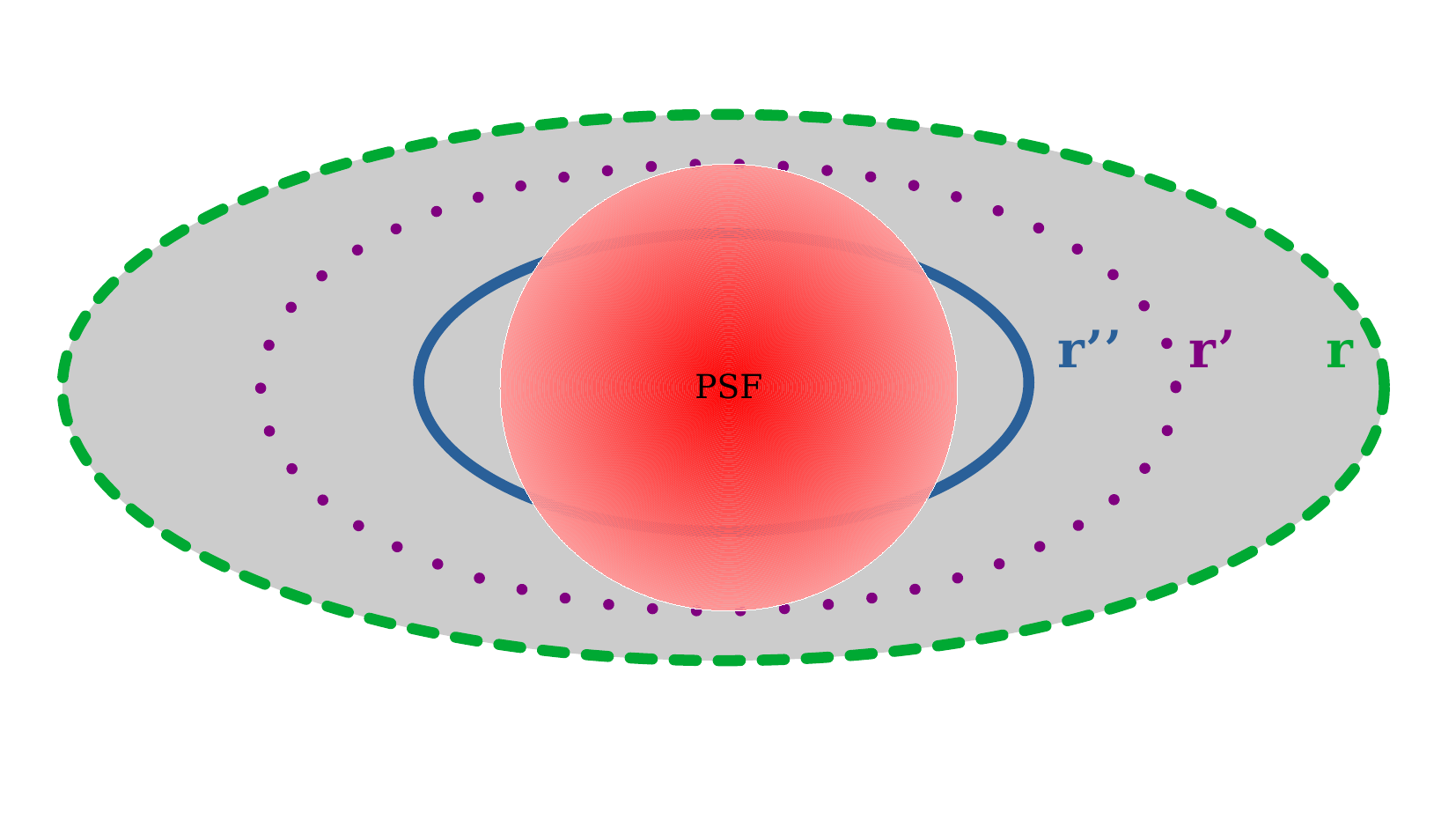}
    \caption{Illustration to show how the shear estimators with different radial weightings are limited by the size of the galaxy and the point spread function (red central circle; PSF). A galaxy with a maximum shear measurement radius $r^{\prime}$ (purple dotted) would be suitable for a single shear measurement, but not for use with the MEM, as any smaller radial weighting - such as $r^{\prime\prime}$ (blue solid) - would be smaller than the PSF. Therefore, a galaxy with a larger maximum measurement radius is required, such as $r$ (green dashed), so that the radial weighting of the second estimator (e.g. $r^{\prime}$) still has a meaningful physical interpretation.}
    \label{fig:radial_scales}
\end{figure}

In order for the different radial weightings of the two estimators to have a meaningful physical interpretation, therefore capturing a difference in the IA amplitude, it is required that the survey in question is able to resolve the physical scales of both radial weightings. An illustration of this issue is shown in Figure \ref{fig:radial_scales}; the galaxy has to be larger than the point spread function (PSF) by a greater degree than would be traditionally required with a single estimator, so that the radial sensitivity of a second estimator can peak at a smaller radius while still probing scales larger than the PSF.

This therefore implies the galaxy sample used with the MEM has stricter requirements on the effective size of a galaxy than a typical lensing sample. Following the formalism of~\cite{Chang2013}, the effective size is given by,
\begin{equation}
    R = \frac{r_{\rm gal}^2}{r_{\rm PSF}^2},
\end{equation}
where $r_{\rm gal}$ is an estimate of the galaxy radius and $r_{\rm PSF}$ is the radius of the point spread function for the galaxy in question. To determine if a galaxy is suitable for shear estimation, the measurement noise, $\sigma_m$, is calculated from,
\begin{equation}
    \sigma_m(\nu, R) = \frac{a}{\nu}\left[1 + \left(\frac{b}{R}\right)^c \right],
\end{equation}
where $\nu$ is the signal-to-noise ratio of the galaxy image and 
$(a,b,c)$ are parameters specific to the shear estimator in consideration. Clearly, a galaxy with a small effective size could still be selected for shear estimation based on high signal-to-noise. Therefore, a cut on the effective size is necessary before determining the measurement noise on the remaining galaxies. This requirement on the effective size is also dependent on the exact radial weighting schemes used. We will address this issue again in section \ref{sec:forecast_data}, in the context of LSST.

\section{Observational Case Study: DES Y1}
\label{sec:case_study}

We now carry out the first application of the MEM to observational data, to probe the effectiveness of the method in the context of a recent weak lensing shape sample. Our primary objectives here are to provide a baseline procedure for applying the method to observational data, and highlight complications that may arise when applying the MEM to real data.
\subsection{Data and shear estimators}
\label{subsec:data}
We choose to use the DES Y1 lens and source catalogues, as there are two different shear estimators applied to the source sample, alongside a multitude of published work that has used these catalogues. This avoids the need to match galaxies between different surveys which have used different estimators, and provides resources for validation and comparison of intermediate measurements~\citep{Prat_2018}.

\subsubsection{Shape catalogues}
\label{sec:des_shapes}

We make use of the two public DES Y1 galaxy shape catalogues~\citep{Zuntz2018}, which assign shears using the \texttt{METACALIBRATION} estimator~\citep[`\texttt{MCAL}' hereafter]{Sheldon2017} and the \texttt{IM3SHAPE} estimator (`\texttt{IM3}' hereafter). The former contains $34.8\times10^{6}$ galaxies, and the latter $21.9\times10^{6}$ galaxies. As previously mentioned, the MEM requires the lensing signal in both estimators to be the same for the lensing shear to cancel. This can naively (neglecting for the moment the complications discussed in Sections \ref{sec:mbias_formalism} and \ref{sec:weights_formalsim}) be achieved by simply selecting only those galaxies for which shear estimates exist from both \texttt{MCAL} and \texttt{IM3}, to create a matched catalogue of galaxies. This matched catalogue contains $17.8\times10^{6}$ source galaxies, with an average effective number density of $n_{\rm s} = 3.26$ arcmin$^{-2}$. Without significant additional analysis, the relative radial sensitivities for these two estimators, and therefore the difference in their IA contamination, is unknown. We therefore proceed with the objective of understanding the intricacies of the MEM in a observational context, and do not necessarily expect a detection of IA.

\subsubsection{Lens catalogue}
\label{sec:des_lenses}

The DES Y1 lens catalogue~\citep{Elvin-Poole2018} contains $660,000$ luminous red galaxy lenses, with redshifts determined via photometry. The \textsc{redMaGiC}~\citep{Rozo_2016} algorithm has been applied to these lenses to select them so as to reduce photo-z error to $\frac{\sigma_{z}}{(1+z)} < 0.02$. For comparative purposes, the photo-z error in the source sample used by~\cite{Blazek2012} was $\frac{\sigma_{z}}{(1+z)} \approx 0.11$. The lens catalogue covers a redshift range of $0.15 \leq z \leq 0.9$~\cite{Elvin-Poole2018} and has a number density of $n_{\rm L}=0.138$ arcmin$^{-2}$.

\subsection{Application methodology}
\label{sec:cs_methods}
Application of the MEM begins with the calculation of the correlation functions required to estimate tangential shear and the boost factor. We make use of the well established \textsc{TreeCorr}\footnote{\url{https://github.com/rmjarvis/TreeCorr}}~\citep{Jarvis2015} package to estimate these quantities which, for DES Y1 shear estimators, are given by,
\begin{align}
    \Tilde{\gamma}^{\rm im3}_{t}(\theta) &=  B(\theta)\left( \frac{\overset{\rm lens}{\underset{j}{\sum}} \Tilde{w}_{{\rm L},j} \Tilde{w}_{{\rm s},j} e_{\rm t}^{j}} {\overset{\rm lens}{\underset{j}{\sum}} \Tilde{w}_{{\rm L},j} \Tilde{w}_{{\rm s},j}(1+m^{\rm im3}_j)} \right),
    \label{eqn:im3_shear}
\end{align}
\begin{align}
    \Tilde{\gamma}^{\rm mcal}_{\rm t}(\theta) &= \frac{B(\theta)}{\langle \boldsymbol{R}_\gamma \rangle + \langle \boldsymbol{R}_{\rm S} \rangle} \left( \frac{\overset{\rm lens}{\underset{j}{\sum}} \Tilde{w}_{{\rm L},j} e_{\rm t}^{j}} {\overset{\rm lens}{\underset{j}{\sum}} \Tilde{w}_{{\rm L},j}} \right),
    \label{eqn:mcal_shear}
\end{align}
where $\Tilde{w}_{{\rm L},j}$ denotes the lens galaxy weights, $\Tilde{w}_{{\rm s},j}$ denotes the source galaxy weights, and $e^j_{\rm t}$ the galaxy ellipticity estimates for each estimator. $\langle R_\gamma \rangle$ is the average of the galaxies' responses to artificial shear, and $\langle R_{\rm s} \rangle$ is an additional response that accounts for selection bias when cuts are made to the catalogue. Notice here that \texttt{MCAL} tangential shear does not incorporate individual galaxy weights, as weighting is implicit in a galaxy's response to artificial shear, hence why the sum is normalised by the average response. We use $10$ angular separation bins log-spaced between $2.5$' and $250$', matching the angular separation range used in~\cite{Prat_2018}, but reducing the number of bins from $20$ to $10$ for greater statistical power in each bin, due to the lower effective source density of the matched catalogue and small magnitude of the IA signal.

To obtain $F$, we do not consider the separations of all random-source pairs individually (as this would be computationally prohibitive). Instead, we split sources and randoms into narrow, weighted redshift bins and count the number of random-source pairs in only those bin combinations within $100$ Mpc/h line-of-sight separation, as well as the total number of random-source pairs in all bins (see equation \ref{eqn:6}).

Finally, in order to obtain the covariance on the measurement, we use a jackknife method with $20$ patches defined by a k-means algorithm (see~\citealp{Jarvis2015} for more detail). The jackknife method can be mathematically expressed via,
\begin{equation}
    \tilde{C}_{\rm JK} = \frac{N_{\rm patch}-1}{N_{\rm patch}}\underset{i}{\sum}(x_{i}-\Bar{x})^{T}(x_{i}-\Bar{x}),
    \label{eqn:cov}
\end{equation}
where the subscript JK denotes that this is a jackknife estimate, $x_{i}$ represents an estimate obtained with patch $i$ excluded, and $\Bar{x}$ represents the average of the $x_{i}$ values. In this case, $x = (1-a)\bar{\gamma}_{\rm IA}$.

We use the entire lens sample to maximise statistical power and ensure a large overlap between the lens and source sample, therefore including as many intrinsically aligned pairs as possible. For future analyses, a narrow lens bin could be preferable to localise the measurement in redshift space for easier comparison to other measurements. However, for the purpose of this case study as a first application of the MEM, maximising the signal-to-noise ratio by including as many galaxy pairs as possible was deemed the appropriate choice. 

\subsubsection{Selection response in the MCAL catalogue}
\label{sec:response}

As discussed previously, if any selection is made to the full \texttt{MCAL} catalogue, a selection response must be calculated to re-calibrate the sample weighting in light of this selection. In the case where no change has been made to the source ellipticity distribution through this selection, the selection response is zero. In the application to DES Y1, the matched catalogue can be thought of as a selection in the \texttt{MCAL} catalogue, based on selection criteria from the \texttt{IM3} catalogue and vice versa.

Such a selection is highly non-trivial to determine \textit{a posteriori}, as it depends on relations between various parameters which contribute to deciding if a certain shear measurement method can be employed for a given galaxy. Furthermore, shear estimation may have failed on certain galaxies for no obvious reason. 

In \cite{Sheldon2017}, the selection response is given by
\begin{equation}
    \langle \boldsymbol{R}_{\rm s} \rangle = \frac{\langle e_{i} \rangle^{S+} - \langle e_{i} \rangle^{S-}}{\Delta \gamma_j},
    \label{eqn:selection}
\end{equation}
where $\langle e_{i} \rangle^{S+}$ and $\langle e_{i} \rangle^{S-}$ represent the mean $i$th component ellipticities measured from images without artificial shear, but with selections made on parameters measured from images with $j$th component shear $\Delta \gamma$ applied positively and negatively.

\begin{table}
    \centering
    \begin{tabular}{cccc}
        \hline
            & $\langle \boldsymbol{R}_\gamma \rangle$ & $\langle \boldsymbol{R}_{\rm s} \rangle$ \\ \hline
            $R_{11}$ & $0.747$ & $-0.049$ \\
            $R_{22}$ & $0.749$ & $-0.044$ \\
            \hline
    \end{tabular}
    \caption{Shear response and selection response for the MCAL matched catalogue. As in~\protect\cite{Prat_2018}, off diagonal terms are negligible. Selection response values have been determined by using the machine learning classifier described in Section \ref{sec:response} and thus their magnitude should be considered a lower limit on the size of the true selection response.}
    \label{tab:selection}
\end{table}

Machine learning methods have the potential to address the challenges of post-hoc selection response estimation in this case, due to their ability to classify large data-sets based on complex relations between `features' of those data-sets. We can, in theory, train a classifier to act as a proxy for the true selection function, by predicting whether or not a galaxy is present in the matched catalogue based on `features' of that galaxy. To do this, we first assign all galaxies in the full \texttt{MCAL} catalogue a new flag, indicating their presence or absence in the matched catalogue. Using the \textsc{scikit-learn}~\citep{scikit-learn} and \textsc{keras}~\citep{chollet2015} packages, we train a multilayer perceptron neural network (see, e.g.~\citealp{Popescu2009}) to predict the probability of a galaxy being selected for the matched catalogue based on its size, \textit{i}, \textit{r} and \textit{z} band flux, signal-to-noise, and $e_{1}$ and $e_{2}$ ellipticity components.

In the case where we take the (overly-simplistic) stance that a galaxy with a selection probability greater than $0.5$ is classed as present in the matched catalogue, the trained classifier is able to predict matched galaxies in unseen test data with an accuracy of $72\%$. However, in reality, the inclusion of a galaxy in the matched catalogue is not a deterministic processes. Therefore, instead, to find $\langle e_{i} \rangle^{S+}$ and $\langle e_{i} \rangle^{S-}$ we calculate the weighted mean of $e_{i}$ over all galaxies in the full \texttt{MCAL} catalogue, with weights given by the classifier's predicted probability of selection for the matched catalogue based on features measured from artificially sheared images. We are then able to estimate the selection response.

We are unable to place uncertainties on the estimated selection response using this method, as the maximal accuracy of $72\%$ (in the hard-cutoff case) indicates that the features available to us in the catalogue do not fully capture the selection and therefore the estimated value is biased from the truth in some unknown way. However, we can consider this value as a lower bound on the magnitude of the selection response and we expect the sign to be correct. A more detailed discussion on the classifier and the interpretation of this value as a lower bound is given in Appendix \ref{app:machine_learning}.

\begin{figure}
    \centering
    \includegraphics[width=0.45\textwidth]{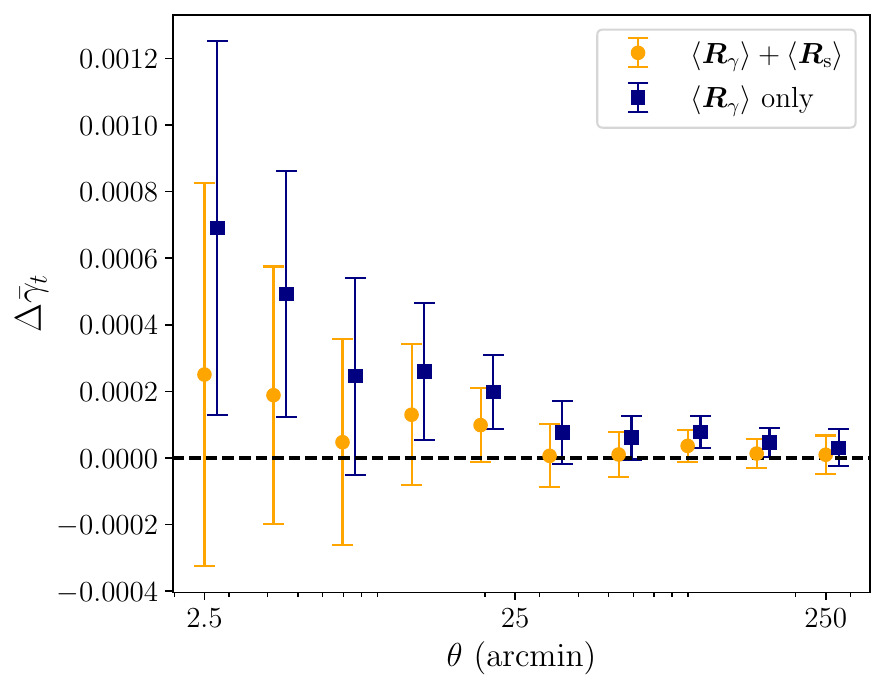}
    \caption{Measurement corrected for selection response: We compare the signal in the cases where selection response is included (yellow circles) and excluded (blue squares). Inclusion of the selection response lower bound gives a measurement which is consistent with zero across all angular separation bins. However, failing to properly account for the selection response leads to a potential false detection across a majority of bins, as shown by the blue data points. Here, and in all following plots, error bars represent $1\sigma (68\%)$ confidence intervals, but are likely underestimated due to the expectation that the true selection response is larger than the value used here. A small offset has been applied on the horizontal axis to improve visual clarity.}
    \label{fig:selection}
\end{figure}

The resulting selection response values and uncertainties are shown in Table \ref{tab:selection}, alongside the shear response values for the matched catalogue. Figure \ref{fig:selection} shows our measurement made using the MEM, both including and excluding the estimated selection response. Failing to include the selection response in this context severely biases the measurement to the point where we see a potential false detection of IA. Importantly, since the value of selection response we have obtained is lower bound, the error bars on the corrected signal should be considered underestimated.

\begin{figure}
    \centering
    \includegraphics[width=0.44\textwidth]{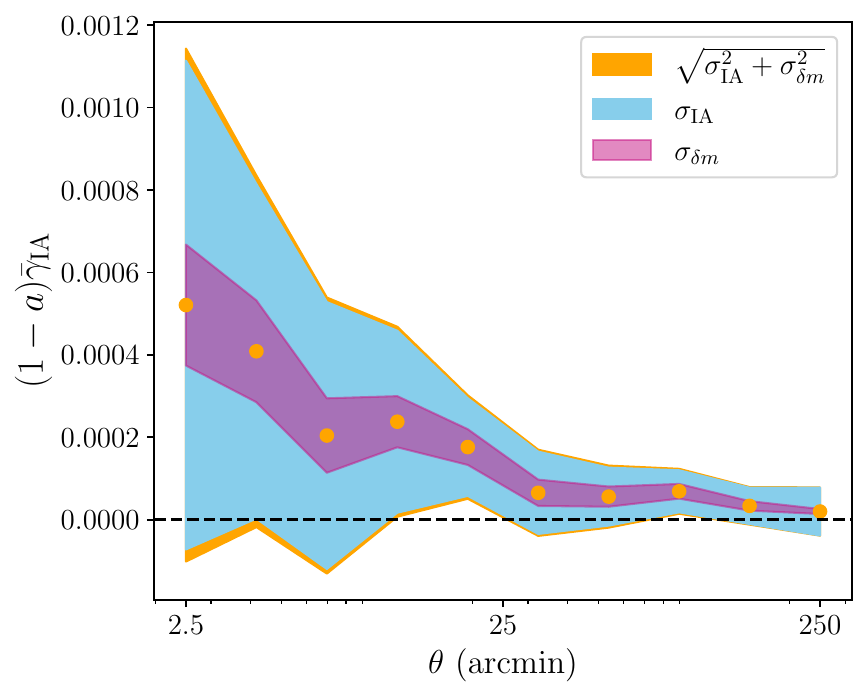}
    \caption{Final $(1-a)\Bar{\gamma}_{\rm IA}$ measurement corrected for selection response, weighting differences, and multiplicative bias uncertainty: Data points are represented by the yellow circles and are the same as in the previous figure. The pink shaded region represents the contribution to $1\sigma$ uncertainty from multiplicative bias uncertainty, while the blue represents the statistical uncertainty obtained from the jackknife method and the uncertainty on the weight induced residual correction. Finally, the orange shows the total $1\sigma$ uncertainty from the combination of both the pink and blue regions. We see that the majority of the known uncertainty in this application of the MEM is statistical.}
    \label{fig:multi_bias}
\end{figure}

We thus present here the first application of the MEM to observational data. Initially, it appears to indicate a null detection of IA when some level of selection response is accounted for. However, as discussed in Section \ref{sec:theory}, there remain potential complications which must be considered before a final measurement is presented. In the figures of the following subsections (Figures \ref{fig:eff_n(z)_signal} and \ref{fig:multi_bias}), as each correction to the signal is applied, the data points and error bars from the previous figure will be represented by navy squares and the corrected signal as yellow circles.

\subsubsection{Galaxy weights and differences in effective source redshift distribution}
\label{sec:effec_red}

In addition to the specific case of the \texttt{MCAL} selection response, more general selection and weighting differences between the \texttt{MCAL} and \texttt{IM3} estimators must be considered. Although using the matched catalogue in the case of \texttt{IM3} and \texttt{MCAL} tangential shear measurements guarantees we use the same literal galaxies for both, the effective contribution to shear from each of these galaxies is different for each estimator. As detailed in \cite{Hoyle2018}, for the DES Y1 case, the effective redshift distribution (which governs the true expected tangential shear) from \texttt{IM3} must account for per-galaxy explicit weights as well as effective weighting by $(1+m_i)$. For \texttt{MCAL}, an effective weighting with respect to the per-galaxy response value is required. Note that the selection response does not factor into this effective weighting, as it is a catalogue level value rather than per-galaxy. One might naively imagine re-weighting the version of the matched catalogue for one estimator by the per-galaxy effective weights of the other estimator, to achieve a unified weighting scheme. However, this is generally ill-advised due to correlations between per-galaxy weights and ellipticities even across different estimators, which could induce a severe bias to our measurement.

Using the two effective weighted redshift distributions, we compute theoretical tangential shears, as seen in equation \ref{eqn:10}, (making use of the Core Cosmology Library, henceforth referred to as CCL;~\citealp{Chisari_2019}) and take their difference to estimate the potential lensing residual, which we will call $\Bar{\gamma}_{\rm L,PA}^{w}$. Provided we trust our theoretical prediction is sufficiently accurate, we can then subtract this from our measured signal to correct for the difference in weighting schemes between the estimators.

In order to ensure we trust our theoretical lensing residual, we consider two primary effects which may affect the accuracy of our predicted $\Bar{\gamma}_{\rm L,PA}^{w}$. Firstly, the fact that the true cosmology is unknown (which also impacts the estimate of galaxy bias used in our modelling) and secondly, un-modelled photometric redshift error.

To account for uncertainty in the true cosmology, we compute $\Bar{\gamma}_{\rm L,PA}^{w}$ for six $\Lambda$CDM cosmologies (parameter sets 3-8 of Table II of~\cite{Abbott2018}) and the $1\sigma$ upper and lower limits on galaxy bias for our sample (estimated from the measurements in~\cite{Elvin-Poole2018} to be $b_{g} = 1.68\pm0.13$). The difference between the smallest and largest lensing residuals for different cosmologies, also incorporating the $1\sigma$ limits on galaxy bias, then gives us an estimate of the uncertainty arising from the true cosmology not being precisely known. To simulate un-modelled photometric error, we convolve the weighted source redshift distributions with a Gaussian photometric error model defined by $\sigma_{z} = 0.1(1+z)$ (which we select to be representative of a worst-case un-modelled uncertainty in a Stage III dataset). Taking the DES \texttt{3x2pt} best-fit parameters to be our fiducial cosmology, the difference between the $\Bar{\gamma}_{\rm L,PA}^{w}$ residuals including and excluding additional photometric error gives an estimate of the uncertainty due to a potential un-modelled photometric error in the data.

\begin{figure}
    \centering
    \includegraphics[width=0.455\textwidth]{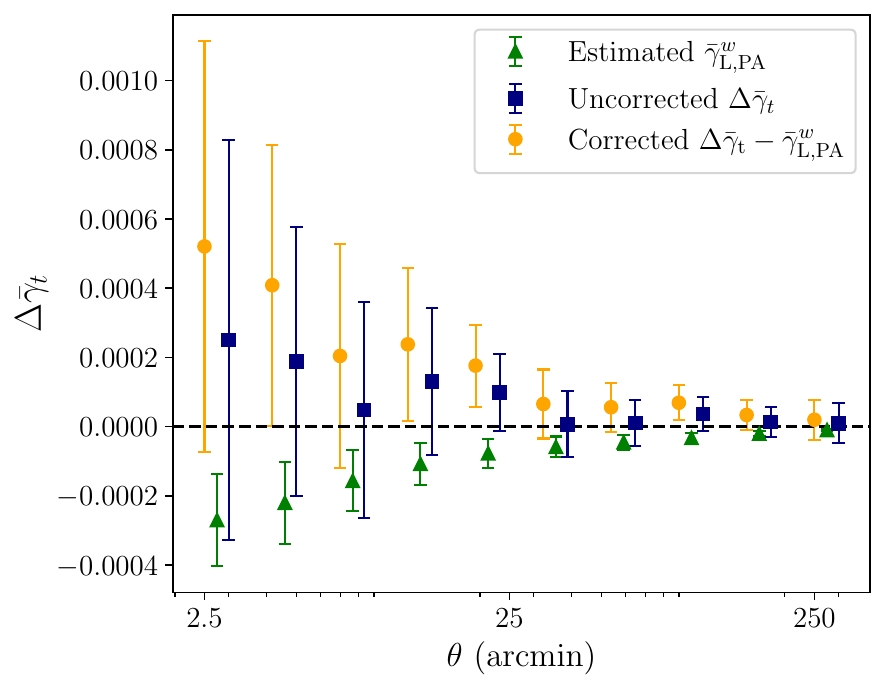}
    \caption{Measurement corrected for selection response and weighting differences: The measured signal minus the weight-induced residual is shown by the yellow circles. Green triangles show the theoretical prediction for this residual, while the blue squares show the signal corrected for the selection response (yellow circles in previous figure), but not for the difference in weights. We see that while the weight induced residual is significant enough to noticeably shift the data points, we still see a null detection in most bins. A horizontal offset is applied to the data points for visual clarity.}
    \label{fig:eff_n(z)_signal}
\end{figure}

We then add these two uncertainties in quadrature to estimate the overall uncertainty on the residual. Figure \ref{fig:eff_n(z)_signal} shows the resulting weight-induced lensing residual, $\Bar{\gamma}_{\rm L,PA}^{w}$, alongside the measured signal with and without $\Bar{\gamma}_{\rm L,PA}^{w}$ subtracted. Error bars on the corrected measured signal have been adjusted to also include uncertainty on $\Bar{\gamma}_{\rm L,PA}^{w}$.

Given the lensing signal is well understood at these scales, it is possible to model and correct for this residual as we have done here. However, in an ideal scenario, a matched weighting scheme would be constructed for both estimators from the outset to avoid this issue entirely.

\subsubsection{Multiplicative bias residuals}
\label{sec:multi_bias}

The final potential contaminant we must address is the multiplicative bias induced lensing residual (detailed in Section \ref{sec:mbias_formalism}). For DES Y1, \texttt{IM3} has a multiplicative bias uncertainty of $m_{\rm im3} = 0.025$ while \texttt{MCAL} has $m_{\rm mcal} = 0.013$. Since we cannot determine the level of multiplicative bias residual in the estimators to greater precision than these associated uncertainties, we choose to treat it as a systematic uncertainty in the MEM and combine it with the estimated covariance matrix. We investigate the significance of the $(m_{\rm im3}-am_{\rm mcal})\Bar{\gamma}_{\rm IA}$ term in equation \ref{eqn:9} and find it to be subdominant compared to the lensing residual term. Since the other complications discussed in this section have been addressed, the measured signal can therefore be expressed as,
\begin{equation}
    \Delta\bar{\gamma}_{\rm t} = \delta m\bar{\gamma}_{\rm L,PA} + (1-a)\bar{\gamma}_{\rm IA},
    \label{eqn:measured_gamma}
\end{equation}
where $\Delta\bar{\gamma}_{\rm t}$ represents the difference in tangential shear between the two shear estimators, which is our measured signal, and $\delta m = (m_{\rm im3}-m_{\rm mcal})$. If we consider the estimated covariance computed from the jackknife method to be the covariance on $\Delta \bar{\gamma}_{\rm t}$, then the covariance for the quantity of interest, $(1-a)\bar{\gamma}_{\rm IA} = \Delta \tilde{\gamma}_{\rm t} - \delta m \tilde{\gamma}_{\rm L, PA}$, is given by:
\begin{multline}
    {\rm Cov}[(1-a)\bar{\gamma}_{\rm IA}(\theta^i) , (1-a)\bar{\gamma}_{\rm IA}(\theta^j)] = \\
    {\rm Cov}[\Delta\bar{\gamma}_{\rm t}(\theta^i),\Delta\bar{\gamma}_{\rm t}(\theta^j)]
    + {\rm Cov}[\delta m\bar{\gamma}_{\rm L,PA}(\theta^i),\delta m\bar{\gamma}_{\rm L,PA}(\theta^j)] \\ 
    - {\rm Cov}[\Delta\bar{\gamma}_{\rm t}(\theta^i),\delta m\bar{\gamma}_{\rm L,PA}(\theta^j)] 
    - {\rm Cov}[\delta m\bar{\gamma}_{\rm L,PA}(\theta^i),\Delta\bar{\gamma}_{\rm t}(\theta^j)].
    \label{eqn:IA_cov}
\end{multline}
Where $i$ and $j$ denote individual $\theta$ bins. To find the lensing residual covariance, we construct a Gaussian distribution for $\delta m$ with a mean $\mu = 0$ and standard deviation $\sigma = \sqrt{m_{\rm im3}^2 + m_{\rm mcal}^2}$. Using $N=1000$ random draws from this distribution and multiplying them by our forecast lensing signal normalised over the boost and F, we calculate the lensing residual covariance using the re-sampling formula,
\begin{equation}
    {\rm Cov}[x^i, x^j] = \frac{1}{N-1}\overset{N}{\underset{k=1}{\sum}}(x_k^i- \bar{x}^i)(x_k^j-\bar{x}^j)^T,
    \label{eqn:resampling}
\end{equation}
where $x_k$ represent individual samples, $\bar{x}$ is the mean value of all samples, $i$ and $j$ again represent $\theta$ bins, and $T$ denotes the transpose. We estimate the cross-covariance between the lensing residual and the measured signal in a similar fashion, by drawing $1000$ samples from a multivariate Gaussian defined by the measured signal and its jackknife covariance, then using the equation \ref{eqn:resampling} to determine the two cross-covariance terms.

At this point we have now, as far as feasibly possible within the scope of this work, corrected or accounted for all the aforementioned complications to the signal. We thus relabel our measurement as the IA component of the MEM, $(1-a)\Bar{\gamma}_{\rm IA}$. Figure \ref{fig:multi_bias} shows our MEM IA measurement data points along with coloured regions representing the various contributions to uncertainty. We see that while residual multiplicative bias does have some noticeable effect on our overall uncertainty, the majority still arises from statistical uncertainty.

\subsection{Summary of findings from DES Y1 case study}

\begin{figure}
    \centering
    \includegraphics[width=0.45\textwidth]{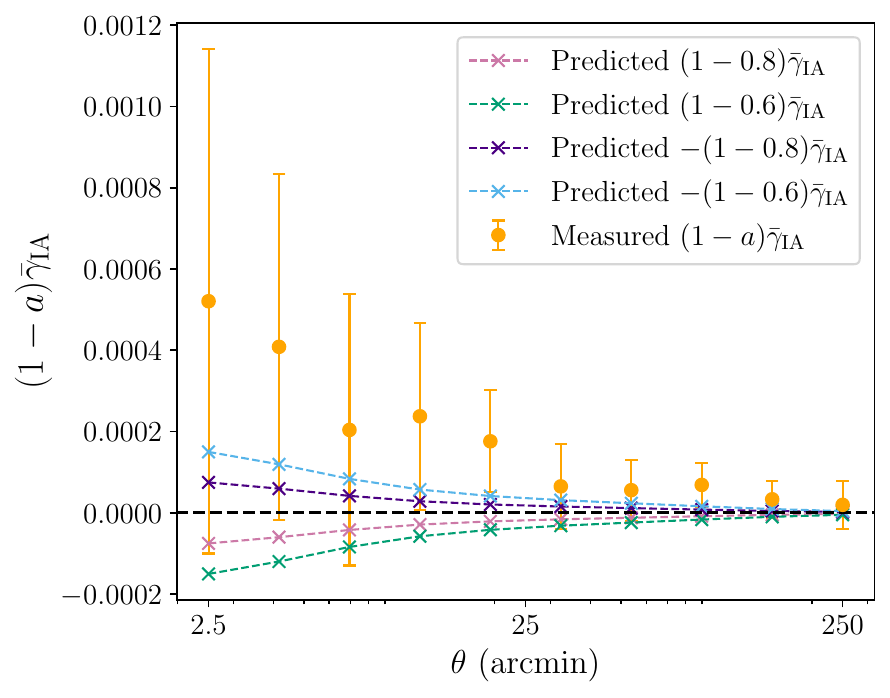}
    \caption{Final MEM measurement of $(1-a)\Bar{\gamma}_{\rm IA}$ compared to the signal expected based on DES Y1 NLA model best fits of $A_{\rm IA}$, for a few example values of $a$. We see our signal, is most consistent with the $-(1-0.6)\Bar{\gamma}_{\rm IA}$ model. We expect the error bars to be underestimated due to the selection response correction representing a lower bound, and that the true selection response would shift the data points closer to zero.}
    \label{fig:des-pred}
\end{figure}

To better contextualise our findings, we compute a theoretical IA signal using CCL and the DES Y1 GGL and clustering best fit non-linear alignment (NLA) model amplitude, $A_{\rm IA} = 0.38$, found in Table 5 of~\cite{Samuroff2019}. Figure \ref{fig:des-pred} shows our final MEM measurement alongside the predicted IA signal for four different hypothetical $a$ values. Given the order of the estimators in our measurement ($\tilde{\gamma}_{\rm t}^{\rm IM3} - \tilde{\gamma}_{\rm t}^{\rm MCAL}$), a positive $(1-a)$ value implies \texttt{IM3} is more sensitive to the outer radial regions of galaxies (because the IA signal is negative in terms of tangential shear), while a negative value, $-(1-a)$, implies the opposite. A $\chi^2$ test shows the strongest agreement with the $-(1-0.6)\Bar{\gamma}_{\rm IA}$ model. We do not consider $a$ values below $0.6$, as this is at the limits of what was seen in \cite{Singh2016}, although from Figure \ref{fig:des-pred} we can expect lower $a$ values in the positive model would give better $\chi^2$ agreement. However, it is imperative we emphasise again that the error bars are likely underestimated, and furthermore, the true selection response is expected to shift the data points closer to zero. We therefore refrain from making any definitive comments on the value of $a$ and the relative radial weightings of the two estimators, and carry out these comparisons simply as an exploratory exercise.

While here we have not been able to detect IA using the MEM, we have made important progress by demonstrating its feasibility for use with observational data. We do not attempt to place any constraints on IA models with this measurement, due to the large uncertainties on our signal, compounded by the expectation that these are underestimated.

To summarise, we will briefly outline the key lessons from this first application of the MEM to observational data, to take forward when considering application of the MEM to Stage IV surveys (the focus of Section \ref{sec:forecasting} and the remainder of this paper).

\begin{itemize}
\item{\textbf{The matched catalogue must be constructed in a way that ensures selection bias can be minimised and / or well-characterised}: Using similar shear estimators could be useful; for example, a modified \texttt{MCAL} estimator with a different radial weighting, used in conjunction with the standard \texttt{MCAL} estimator.}

\item{\textbf{Establishing a shared weighting scheme is highly beneficial to ensure differences in weighting (explicit or effective) do not manifest as differences in the lensing shear}: While differences in weighting can be treated provided we trust our modelling of the lensing signal, future applications should make it a priority to construct a matched weighting scheme for both estimators \textit{a priori} to avoid this process entirely.} Alternatively, another approach could be to forgo weighting galaxies altogether if signal-to-noise allows (see e.g.~\citealp{zhang2023}).

\item{\textbf{Multiplicative bias uncertainty must be demonstrably subdominant or accounted for within the overall measurement uncertainty}: Future applications should select estimators with the lowest levels of calibration uncertainty possible. The uncertainty should also be accounted for in conjunction with the measurement covariance using the formalism introduced in Section \ref{sec:multi_bias}.}
\end{itemize}

\section{Forecasting for stage IV surveys}
\label{sec:forecasting}

Given our findings from the DES Y1 case study, we now present a forecast for the performance of the MEM with synthetic data sets representative of a Stage IV lensing survey. Specifically, we consider LSST Y1 and Y10. For comparison, the relative survey specifications for DES and LSST are given in Table \ref{tab:survey}. Using this forecast, we will seek primarily to address the issue of residual multiplicative bias as done in Section \ref{sec:multi_bias}, by removing the assumption of sub-dominance and instead accounting for it within the measurement uncertainty. This will then allow us to place requirements on shear estimators in two cases: detection of IA with the MEM, and constraint of the IA scale dependence with the MEM. We choose to focus here on multiplicative bias because, as established in Section \ref{sec:case_study}, issues with selection bias and differences of effective weighting schemes are more readily overcome, while multiplicative bias uncertainty must be included as a systematic of the MEM.

\begin{figure*}
    \begin{subfigure}
        \centering
        \includegraphics[width=0.45\textwidth]{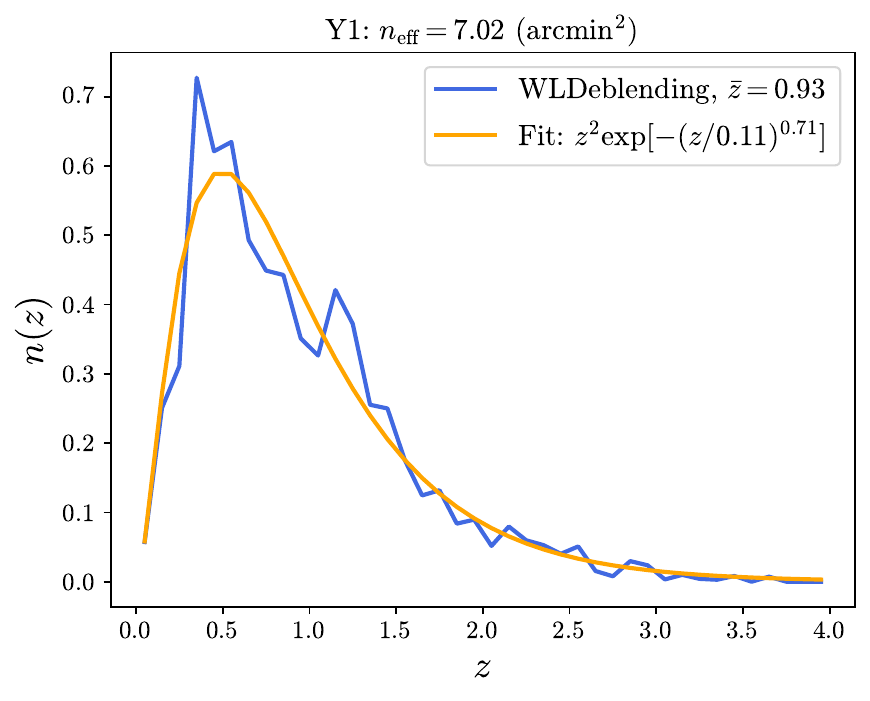}
    \end{subfigure}
    \begin{subfigure}
        \centering
        \includegraphics[width=0.45\textwidth]{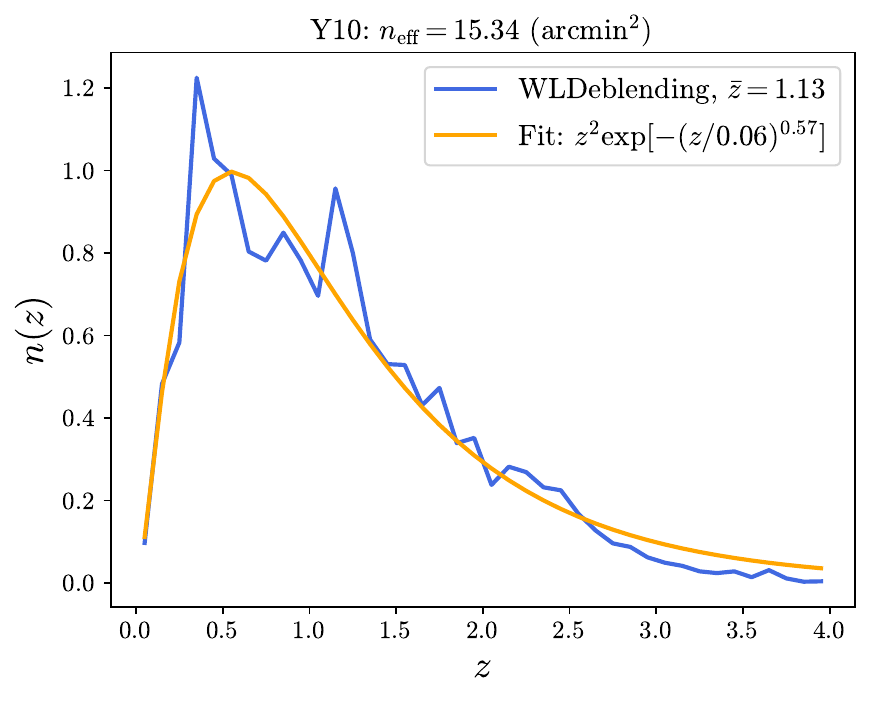}
    \end{subfigure}
    \caption{LSST Y1 (left) and Y10 (right) source galaxy redshift distributions obtained after the effective size cut. The parametric model from equation \ref{eqn:lsst_dndz} has been re-fit to obtain a new model. The $n_{\rm eff}$ values correspond to a loss of roughly $2$ and $11$ sources per square arcminute for the Y1 and Y10 cases respectively, when compared to the original values (see Table \ref{tab:survey}).}
    \label{fig:lsst_neff}
\end{figure*}

For a pair of hypothetical shear estimators, we will forecast MEM performance with respect to their amplitude offset parameter, $a$, and the Pearson correlation coefficient~\citep{cohen2009pearson} of their shape-noise, $\rho$. It is important to note here that we do not seek to place strict limits or requirements on observational choices or shape estimators. Such limits would be specific to the analysis choices made here and the survey in question. Instead, we aim to provide more general guidelines and targets for the development of bespoke shape estimators to be used with the MEM.

\subsection{Preparation of synthetic data vector}
\label{sec:forecast_data}
\subsubsection{Redshift distributions}

In order to carry out the forecasting, we assume the prescriptions for lens and source galaxy samples given in the LSST DESC Science Requirements Document v1 (\citetalias{LSST_SRD};~\citealp{LSST_SRD}). Where sample-dependent parameters are referenced, unless otherwise stated, it can be assumed the associated values are taken from \citetalias{LSST_SRD}.

To address the issue of galaxy size discussed in Section \ref{sec:radial_weight_res}, we impose a strict effective size size cut of $R \geq 3$. Relating this to specific radial weightings and values of $(1-a)$ would require analysis of galaxy images, which is beyond the scope of this work and we defer to a future analysis. However, we anticipate that such a cut would likely be sufficient and potentially even excessive. Determining exact values for a given pair of estimators will be the subject of future analysis.

Given this cut, we re-compute the \citetalias{LSST_SRD} redshift distributions~\footnote{To do this, we use a modified version of this Jupyter notebook: \url{https://github.com/LSSTDESC/Requirements/blob/master/notebooks/RedshiftDistributions.ipynb}} using \textsc{WeakLensingDeblending}~\footnote{\url{https://github.com/LSSTDESC/WeakLensingDeblending/blob/master/docs/index.rst}}~\citep{Sanchez2021} simulated galaxy catalogs for Y1 and Y10, where Y1 is defined as being $10\%$ of the total 10-year exposure time. Figure \ref{fig:lsst_neff} shows the raw distributions and best fit to the \citetalias{LSST_SRD} parametric distribution given by,
\begin{equation}
    \frac{dN}{dz} \propto z^2 {\rm exp}\left[-\left(\frac{z}{z_0}\right)^\alpha\right],
    \label{eqn:lsst_dndz}
\end{equation}
with constants $z_0$ and $\alpha$ defined in \citetalias{LSST_SRD} for the Y1 and Y10 lens galaxy samples, and in Figure \ref{fig:lsst_neff} for the Y1 and Y10 source galaxy samples used here. Compared to the values in Table \ref{tab:survey}, the larger drop in $n_{\rm eff}$ for Y10 is due to the larger fraction of galaxies removed by the effective size cut in Y10, since Y1 will likely image the biggest and brightest galaxies first, with smaller, fainter ones resolved over the next 9 years.

\begin{table}
\centering
\begin{tabular}{lccc}
\hline
            & DES Y1 & LSST Y1 & LSST Y10 \\ \hline
$f_{\rm sky}$ & 0.0321 & 0.436   & 0.436    \\
$n_{\rm eff}$ (arcmin$^{-2}$)        & 6.38   & 9.52    & 26.94    \\
$z_{\rm s}^{\rm max}$        & 1.3    & 4.0     & 4.0      \\
$m_{\rm max}$        & 0.028  & 0.013   & 0.003    \\ \hline
\end{tabular}
\caption{Relevant survey parameters for DES Y1, and LSST Y1 and Y10. $f_{\rm sky}$ is the fraction of the sky that was or will be covered by the survey. We can see LSST promises a significant increase in statistical power over DES Y1. To ensure this greater sample size is utilised effectively, shear estimators will need to demonstrate significantly lower $1\sigma$ limits on multiplicative bias uncertainty, defined as $m_{\rm max}$}. The MEM will also benefit from these decreased levels of multiplicative bias uncertainty.
\label{tab:survey}
\end{table}

We use only one source and one lens tomographic bin, as was done in the DES Y1 application. For lenses, we use a narrower bin compared to the DES Y1 case study, in the range $1.0 \leq z_{\rm l} \leq 1.2$, which represents the highest redshift lens bin defined in \citetalias{LSST_SRD} for Y1. We choose to use this bin so as to consider a hypothetical scenario where we are attempting to probe IA in a narrow slice of redshift space. Choosing the highest redshift bin also results in a lower number of sources far behind the lenses, which will contribute to lensing residuals but not the IA signal. For the source bin, we use the full redshift range between $0.05 \leq z_{\rm s} \leq 3.5$. Placing some additional limit on the maximum source redshift would also prove beneficial to reducing lensing residuals, however, it would be detrimental to the statistical uncertainty on our $\gamma_{t}$ measurements, which we expect to be a crucial limiting factor in our error budget, especially given the findings of Section \ref{sec:case_study}. We also account for photo-z uncertainty where appropriate, by convolving equation \ref{eqn:lsst_dndz} with a Gaussian uncertainty model from \citetalias{LSST_SRD},
\begin{equation}
    p(z_{\rm s},z_{\rm ph}) = \frac{1}{\sqrt{2\pi}\sigma_z}{\rm exp}\left[-\frac{(z_{\rm ph}-z_{\rm s})^2}{2\sigma_z^2}\right],
    \label{eqn:photz_err}
\end{equation}
where $z_{\rm s}$ and $z_{\rm ph}$ represent spectroscopic and photometric redshift respectively and $\sigma_z$ is defined as $0.05(1+z_{\rm s})$for sources and $0.03(1+z_{\rm s})$ for lenses.

It is important to mention that including the full source sample behind the lenses has the potential to increase the magnitude of the lensing signal and thus any lensing residuals. In this work, we will only consider the full sample to try and obtain limits on the acceptable values of $a$ and $\rho$ in the instance where statistical uncertainty requires we use the full sample. However, in a real analysis, some benefit could be gained from placing a lower upper limit on the source redshift bin, to restrict the number of sources behind the lenses which are contributing to the lensing shear, but not the IA shear. The exact limit will depend on the survey in question, as a balance would need to be struck between retaining an acceptable signal-to-noise ratio for the overall shear, whilst minimising the lensing residual.

\subsubsection{Halo Occupation Distributions}

To calculate theoretical tangential shears for lensing and IA, as well as the boost, we first require power spectra for the quantities of interest. To obtain predictions within the 1-halo regime, where we are most interested in using the MEM, we use the halo model formalism~\citep{Seljak2000,Peacock_2000,COORAY_2002}, 
\begin{multline}
       P_{uv}(k) = \int dM n(M) \langle u(k|M) v(k|M)\rangle \\
       + \int dM n(M) b(M) \langle u(k|M)\rangle\ \int dM n(M) b(M) \langle v(k|M)\rangle P_{\rm lin}(k).
       \label{eqn:halo_model_pk}
\end{multline}
Here, $k$ is the wavenumber, $M$ is halo mass, $P_{\rm lin}(k)$ is the linear power spectrum, $n(M)$ is the number of halos of mass $M$, $b(M)$ is the halo bias as a function of halo mass~\citep{Tinker_2010}, and $u(k|M)$ and $v(k|M)$ are the Fourier space halo profiles of the tracers being correlated. The first and second terms of equation \ref{eqn:halo_model_pk} represent the 1-halo and 2-halo contributions respectively.

In the context of this work, the profiles needed are; lens galaxy density, given by the halo occupation distribution (HOD) of~\cite{Nicola_2020}; source galaxy density, given by the HOD of~\cite{Zu_2015}; matter density, given by the Navarro-Frenk-White profile~\citep{Navarro_1996,Navarro1997}; and a satellite shear HOD for intrinsic alignment in the 1-halo regime~\citep{Schneider_2010,Fortuna2021}. The lens HOD was chosen as it is based upon Hyper Suprime Cam data and therefore able to model a sample with a number density somewhat representative of deep, LSST observations. Similarly, the source HOD can be modified using the source number density for our LSST-like sample. The satellite shear HOD is capable of modelling 1-halo IA effects, making it an ideal candidate for constructing a theoretical IA signal in this context. In Appendix \ref{app:power}, we verify our implementations of the 2-point cumulants in equation \ref{eqn:halo_model_pk} are correct.

From the halo model and the above HODs, we obtain the 1-halo and 2-halo terms for lensing shear and the boost. For IA, we only obtain the 1-halo term from the halo model, the 2-halo IA term instead comes from an NLA model \citep{Bridle_2007}. We note that this does mean there is some discrepancy between the modelling of our lensing and IA signals. In an ideal scenario, the full IA signal would have been modelled using the halo model. However, as discussed in Appendix B of \cite{Fortuna2021}, the IA halo model is currently unable to accurately describe the transition scales, particularly in GGL, and determining the correct description is a significant study itself which is beyond the scope of this work. We therefore follow their approach by using the NLA model to compensate for the lack of power that results from this incomplete description. We do not attempt a similar treatment for the lensing signal (which may also lack power in these transition scales as evidenced by~\citealp{Mead2021,Mahony2022}), as we do not expect it to meaningfully impact our forecasting results. Having introduced our fundamental modelling choices, we will now go on to describe the modelling procedure in greater detail.

\subsubsection{Lensing and IA shears}

For on-sky separation binning, we consider seven log-spaced bins in the projected separation range $0.1 \leq r_{\rm p} \leq 10$. Unlike in DES Y1, here, we choose to consider $r_{\rm p}$ instead of angular separation $\theta$, for easier comparison of this work to \citetalias{Leonard2018} and~\cite{Blazek2012}. This choice means we are accessing the 1-halo regime, where we are most interested in using the MEM to study IA scale dependence.

As in Section \ref{sec:case_study} above, the theoretical lensing shear is obtained using CCL~\citep{Chisari_2019}, but this time using the halo model to compute both the 1-halo and 2-halo contributions to the galaxy-matter power spectrum, $P_{\rm gM}(k)$. The power spectrum can then be used to obtain the angular power spectrum using the Limber approximation,
\begin{equation}
    C_{\rm gM}^{\rm len}(\ell) = \int d\chi \frac{W_{g}(\chi)W_{l}(\chi)}{\chi^2}P_{\rm gM}\Bigg(\frac{\ell + 1/2}{\chi}, z(\chi)\Bigg),
    \label{eqn:lensing_aps}
\end{equation}
where $\ell$ is the angular multipole at which the spectrum is defined, $\chi$ is comoving line-of-sight distance (which is a function of redshift), and g and M refer to galaxies and matter respectively. $W_{g}(\chi)$ is the galaxy density function given by,
\begin{equation}
    W_{g}(\chi) = n_g(z) \frac{dz}{d\chi}.
\end{equation}
For the purpose of this theoretical modelling, we will redefine the lensing efficiency function (previously given by equation \ref{eqn:lens_effic}) as $W_l(\chi)$,
\begin{equation}
    W_l(\chi) = \frac{3H_{0}^{2}\Omega_{m}}{2c^2}\frac{\chi}{a(\chi)} \int d\chi^{\prime} n(z(\chi^{\prime}))
    \frac{dz}{d\chi^\prime}\frac{\chi^\prime-\chi}{\chi^\prime}.
\end{equation}
Here, $\Omega_m$ is the matter fraction of the Universe, and $H_0$ is the Hubble constant, $a(\chi)$ is the scale factor and $c$ is the speed of light. Note that we take the fiducial cosmology to be the same as \citetalias{LSST_SRD} and assume a flat Universe ($\Omega_k=0$). Assuming B-modes are zero, we can then find the projected correlation function in real space (which is analogous to the tangential shear in this context) using,
\begin{equation}
    \gamma_{\rm t}(\theta) = \underset{\ell}{\sum} \frac{2\ell + 1}{4\pi} C^{\rm len}_{\rm gM}(\ell) d^{\ell}_{0,2}(\theta),
\end{equation}
where $\theta$ is the angular separation of the tracers in question and $d^\ell_{0,2}$ are the Wigner-d matrices for tracers with spins $0$ (galaxies) and $2$ (shear). We solve this equation via a brute force sum to avoid instabilities that arise when considering the high $\ell$ values required to probe the 1-halo regime.

To model the theoretical IA signal, we draw on several areas of the literature. We compute the lens position and 1-halo satellite alignment angular cross spectrum, $C^{1\rm{h}}_{\rm gI}(\ell)$ from the prescription of~\cite{Fortuna2021} (hereafter: \citetalias{Fortuna2021}). To account for the different red and blue galaxy alignment amplitudes, we take a weighted average of the 1-halo amplitudes for red and blue galaxies given in \citetalias{Fortuna2021}. For LSST Y1, we use an approximate red fraction of $f_{\rm red} = 0.10$, which decreases to $f_{\rm red}=0.05$ for Y10. We note that these values are not rigorously determined, but rather qualitative estimates. We vary the value of $f_{\rm red}$ within a reasonable range of $0.00 \leq f_{\rm red} \leq 0.30$ and find a negligible effect on the 1-halo signal, due to the small difference between the red and blue galaxy 1-halo amplitudes found in~\citetalias{Fortuna2021}. We also take the projected separation scale dependence parameter, $b=-2$, from~\citetalias{Fortuna2021}.

A more significant effect could arise from differences in the luminosity functions of~\citetalias{Fortuna2021} and LSST. Here, a detailed analysis of these luminosity functions is not appropriate, so we instead determine lower limits on the average alignment amplitude, below which the maximum lensing residual dominates the 1-halo IA signal. These are $a_{\rm 1h} > 1.00\times10^{-4}$ and $a_{\rm 1h} > 5.00\times10^{-5}$ for Y1 and Y10 respectively.

We also compute a second angular cross spectrum for the 2-halo regime, $C^{\rm NLA}_{\rm gI}(\ell)$, using a redshift dependent NLA model~\citep{Hirata_2007,Bridle_2007} following equation 24 of \cite{Secco2022}, with best fit parameters taken from row three of Table III in \citep{Secco2022} (lens bias is also accounted for using the prescription given in \citetalias{LSST_SRD}). 

It is important to note that our choice of fiducial IA model may not represent what is eventually seen in LSST data, due to the increased depth of LSST compared to the samples used in~\citetalias{Fortuna2021} and DES Y3. However, it is nonetheless necessary for us to choose some fiducial IA model in the absence of any measurements truly representative of LSST. We refer the reader to~\cite{Krause2016} for discussion on the complexities of forecasting the IA contamination to stage IV surveys. We caution that the findings of this study may not apply if the true signal is found to be vastly different to the models we have used here, but ultimately it is necessary for us to make some modelling choices. We defer a detailed comparison of different IA models to future work.

\begin{figure}
    \centering
    \includegraphics[width=0.45\textwidth]{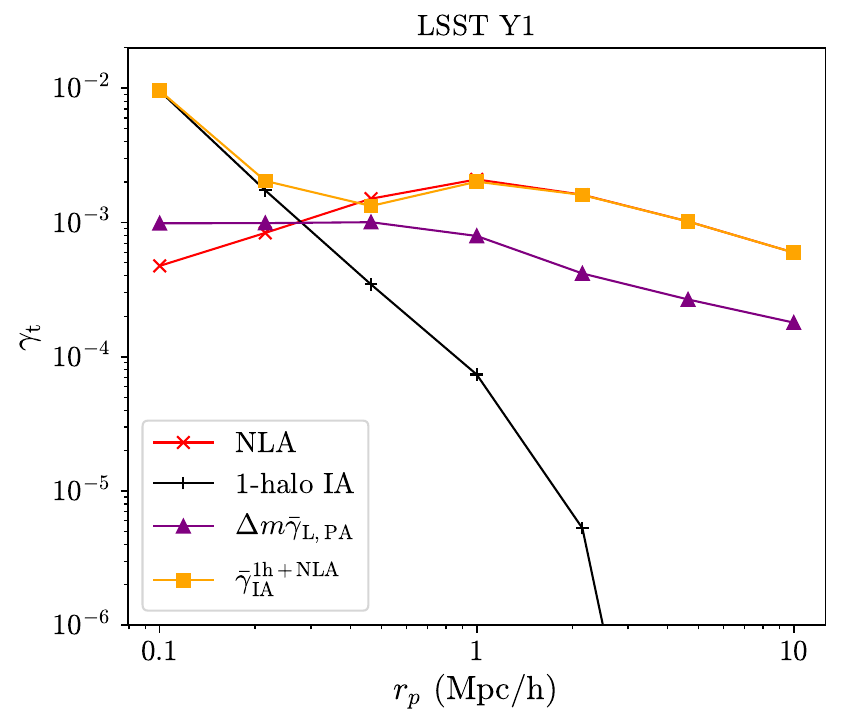}
    \caption{Comparison of tangential shears for the different IA models discussed. The windowed tangential shear (yellow squares) is the model used in our forecasting. It is obtained via a truncated combination of the 1-halo (black crosses) and NLA (red crosses) models. The maximum lensing residual (purple triangles) is shown for comparison. We show the absolute value of the IA signal for easier visual comparison to the lensing residual.}
    \label{fig:gammaT}
\end{figure}

Having obtained the 1-halo and 2-halo spectra, we then combine them, truncating each via a window function to avoid double counting in the 1-halo to 2-halo transition,
\begin{equation}
    C^{\rm 1h+NLA}_{\rm gI}(\ell) = C^{\rm 1h}_{\rm gI}\left(1-{\rm exp}\left[\left(-\frac{\ell}{\ell_{\rm1h}}\right)^2\right]\right) + C^{\rm NLA}_{\rm gI}\left({\rm exp}\left[\left(-\frac{\ell}{\ell_{\rm2h}}\right)^2\right]\right),
    \label{eqn:c_ell}
\end{equation}
with $\ell_{\rm1h} = 1.4\times10^4$ and $\ell_{\rm2h}=3\times10^4$ chosen to give as smooth a transition as possible, and subscript I denoting an intrinsic shape tracer. Table \ref{tab:ia_params} contains a summary of the key IA model parameters used in this work. We then use this angular power spectrum to compute the IA tangential shear for our choice of projected and tomographic bins. Figure \ref{fig:gammaT} shows the various tangential shears discussed here. All quantities shown have been normalised by the estimated number of physically associated pairs, which is why we see a flattening of the lensing signal in the scales where the boost factor strongly dominates and contribution from $F$ is negligible, as the boost factor has a similar scale dependence to the lensing signal.

\begin{table}
\centering
\begin{tabular}{lcccc}
\hline
           & $A_{\rm red}$ & $A_{\rm blue}$ & $b$    & $\ell_{\rm 1h}$ \\
           &            &           &       & \\
1-halo     & $0.001$  & $0.0006$  & $-2$   & $1.4x10^4$ \\ \hline
           & $A$      & $z_{0}$    & $\eta$  & $\ell_{\rm 2h}$ \\
           &            &           &           & \\
NLA 2-halo & $0.36$   & $0.62$    & $1.66$ & $3x10^4$   \\ \hline
\end{tabular}
\label{tab:ia_params}
\caption{Key parameters and their values used in our IA modelling. For the 1-halo term, $f_{\rm red}$ values of $0.10$ and $0.05$ were used for LSST Y1 and Y10 respectively. 1-halo values are taken from~\protect\citetalias{Fortuna2021}, while the NLA 2-halo values are from~\protect\cite{Secco2022}.}
\end{table}

\subsubsection{Galaxy weights and a new definition for F}

In keeping with \citetalias{LSST_SRD}, we adopt a simple weighting scheme for all source galaxies given by,
\begin{equation}
    \Tilde{w}_j = \frac{1}{\sigma^2_\gamma + (\sigma_e^j)^2},
\end{equation}
with intrinsic shape noise, $\sigma_\gamma = 0.12$ and per component error on the measured ellipticity of a given galaxy $j$, $\sigma_e^j = 0.26$, such that all galaxies are weighted equally. We also adopt a new definition for $F$~\citep{Blazek2015,Safari2023}, by redefining the maximum line-of-sight separation at which we expect galaxies to be physically associated and therefore contributing to the average IA signal in a particular projected separation bin,
\begin{equation}
    \Pi(r_{\rm p}) = 
    \begin{cases}
        r_{\rm p} & \text{if } r_{\rm p} > 2 \\
        2 & \text{if } r_{\rm p} \leq 2
    \end{cases}.
    \label{eqn:f_theta}
\end{equation}
This definition of $F$ extends on the previous one to remove the contribution of pairs that are close in projected separation, but far in line-of-sight separation. The lower limit of $2$ Mpc/h ensures that all galaxies within the same halo are included, regardless of how small their projected separation may be. We note that if this definition is used with large projected separations, some upper limit on the line-of-sight separation may be required to avoid artificially diluting the signal. However, this will not pose a problem in our analysis, as we do not consider $r_{\rm p}$ beyond $10$ Mpc/h. More detail on the calculation of $F$ and the boost factors is given in Appendix \ref{app:boost}.

\subsection{Residual multiplicative bias in LSST-like estimators}
\label{sec:lsst_mbias}

Because we are not looking to cross correlate different source tomographic bins, we assume a constant multiplicative bias across all redshifts. In \citetalias{LSST_SRD}, the allowed levels of multiplicative bias uncertainty given are $\pm0.013$ and $\pm0.003$ for Y1 and Y10 respectively.

\subsubsection{Forecasting procedure}

We use the \textsc{TJPCov}~\footnote{\url{https://github.com/LSSTDESC/TJPCov}} package to estimate the statistical covariance on our forecasts, and validate our approach against the \citetalias{LSST_SRD} forecast covariance. It is important to note that inclusion of a residual multiplicative bias alters the covariance expression given in \citetalias{Leonard2018}; we address this in Appendix \ref{app:cov}, but find it amounts to only percent level corrections to the original expression, so therefore compute the statistical covariance in the same way as \citetalias{Leonard2018}. 

The maximum allowed multiplicative bias calibration uncertainties from \citetalias{LSST_SRD} are
$m = 0.013$ and $m = 0.003$ for Y1 and Y10 respectively. Using the same process detailed in Section \ref{sec:multi_bias}, we estimate the IA only covariance by re-sampling from a Gaussian distribution of multiplicative bias uncertainty values. In this case, to estimate $\Delta \Bar{\gamma}_t$ for the cross covariance terms, we simply add our fiducial IA signal and the mean lensing residual and use the \textsc{TJPCov} statistical covariance to define the multivariate Gaussian. We again estimate the IA component of the residual multiplicative bias, $(m-am^\prime)\Bar{\gamma}_{\rm IA}$, but find it is only significant when $a$ is very large. In this case, the lensing residual would dominate the IA signal, and any future applications (by nature of the method) should seek to minimise $a$. We therefore assume this term to be negligible.

\begin{figure*}
    \begin{subfigure}
        \centering
        \includegraphics[width=0.47\textwidth]{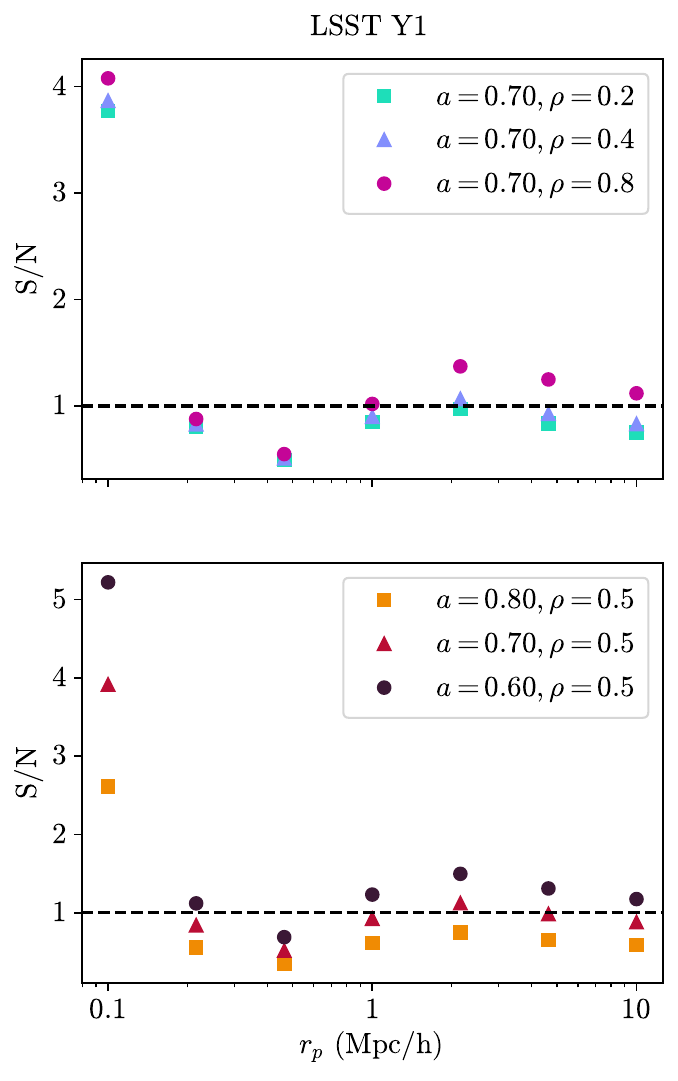}
    \end{subfigure}
    \begin{subfigure}
        \centering
        \includegraphics[width=0.48\textwidth]{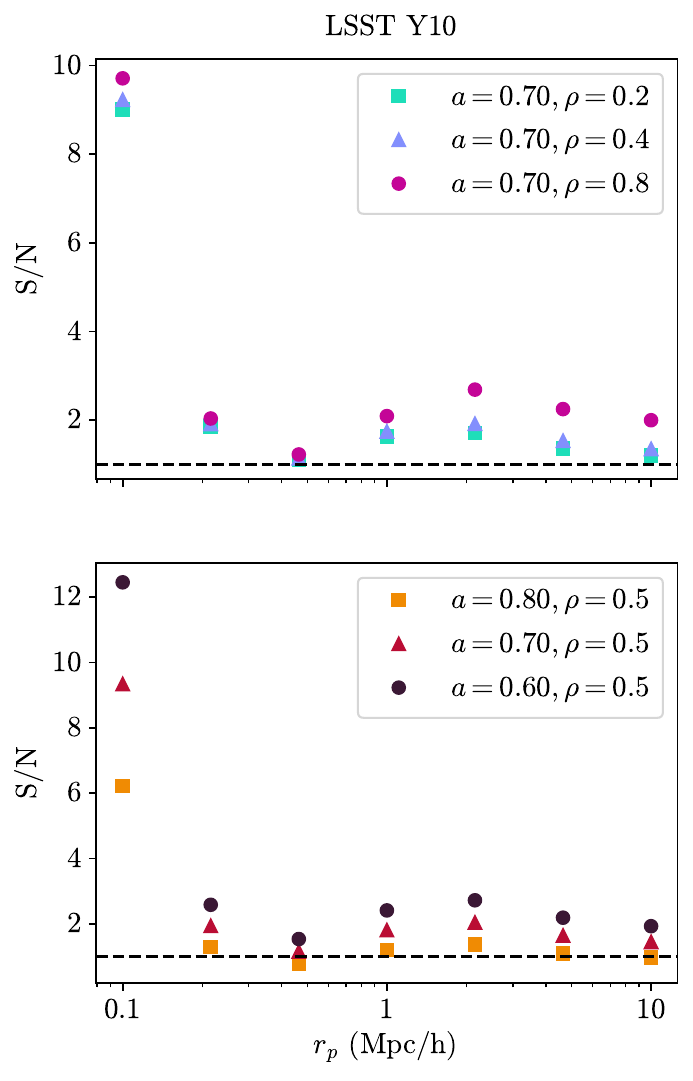}
    \end{subfigure}
    \caption{SNR in each $r_{\rm p}$ bin taken as the forceast IA signal divided by the $1\sigma$ uncertainty. Values greater than 1 represent a detection of IA above statistical noise and uncertainty due to multiplicative bias, for the given $r_p$ bin in isolation. LSST Y1 is shown on the left and Y10 on the right. The top panels show different cases where $\rho$ is varied but $a$ kept fixed, while the bottom panels show the opposite. A much higher SNR is seen in the lowest projected separation bin where the 1-halo term becomes highly dominant. Varying $a$ has a more significant effect on the SNR than varying $\rho$. However, in the lowest signal to noise bins $\rho$ can be the difference between a detection and a signal consistent with zero. In Y10 compared to Y1 we see a roughly factor of $2$ increase in per-bin signal to noise.}
    \label{fig:snr_diag}
\end{figure*}

In this forecasting scenario, we are capable of isolating the IA component and measuring the signal-to-noise, and so for the following results we will focus on this to better understand the region in the $a$-$\rho$ parameter space we should ideally target. While this is not the case in an observational scenario, because the mean of the residual multiplicative bias distribution should be zero, in most cases, the sample mean of the lensing residuals is much smaller than the IA signal, and equation \ref{eqn:measured_gamma} is dominated by the IA component. Therefore, as shown in Section \ref{sec:multi_bias}, in an observational context following this procedure is still appropriate, as it widens the uncertainty on the signal to account for residual bias larger than the expected value. Even if the measured signal were to include a significant contribution from a lensing residual, this would be captured by large cross-covariance between the residual and the measured signal.

\subsection{Forecasting results}

\begin{figure*}
    \begin{subfigure}
        \centering
        \includegraphics[width=0.48\textwidth]{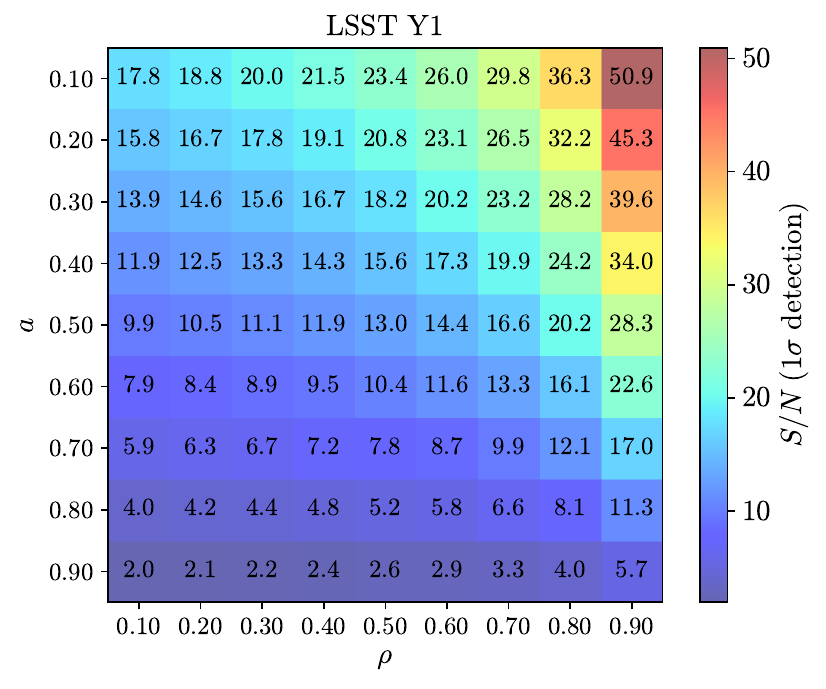}
    \end{subfigure}
    \begin{subfigure}
        \centering
        \includegraphics[width=0.48\textwidth]{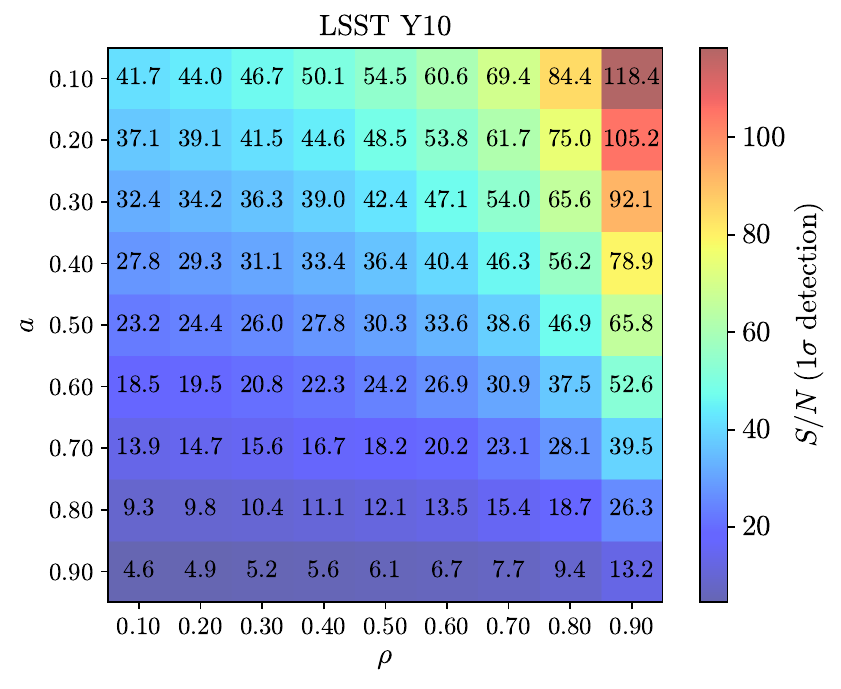}
    \end{subfigure}
    \caption{Combined SNR in all $r_{\rm p}$ bins across the $a$-$\rho$ parameter space for LSST Y1 (left) and LSST Y10 (right). We see high signal to noise in the entire explored region, even in areas where we expect the majority of the signal to have $1\sigma$ uncertainty consistent with zero. This is a result of the high SNR in the lowest separation bin and highly correlated off diagonal elements of the covariance matrix.}
    \label{fig:overall_snr}
\end{figure*}

The key variables of the method which can potentially be tuned and controlled are the amplitude offset parameter, $a$, and the shape noise correlation between the estimators, $\rho$. We therefore choose to forecast the signal-to-noise ratio (SNR) for different values of $\rho$ and $a$, to place requirements on these parameters for an IA signal to be detected in LSST data, whilst accounting for realistic levels of residual multiplicative bias.

We will consider three different definitions of the SNR to better contextualise our results: the SNR in a single $r_{\rm p}$ bin, the combined SNR for all $r_{\rm p}$ bins including the full covariance matrix, and the SNR for a 1-halo scale dependence parameter fit using a Markov-Chain Monte-Carlo (MCMC) method.

\subsubsection{The amplitude offset parameter, $a$}
\label{sec:a}

In order to better contextualise our results, here we will briefly discuss $a$, its range of expected values from observational results, and potential complications that may arise when attempting to determine its value.

In an observational context, without prior knowledge on the radial sensitivity of the two estimators, it is difficult to determine precisely the value of $a$. It could be possible to gain further information on $a$ through the inclusion of a third shear estimator, as this would allow for the ratio of different combinations of shear estimators to be compared. However, it is likely the additional computational cost and necessary preparation for including a third shear estimator would be prohibitive and therefore we do not carry out any investigation into this. Regardless, even if $a$ cannot not be estimated, a constraint on the scale dependence alone would allow for the amplitude of an IA model to be left free and calibrated within cosmological parameter estimation pipelines. Therefore, it is not essential to the success of the MEM that $a$ be accurately estimated.

Despite this, in the context of designing optimised MEM estimators, it is important to have a rough target value for $a$, to ensure constraining the scale dependence is possible. Radial sensitivity calibration for the estimators could be carried out on simulated images, where the dependence of the IA signal on galaxy scale can be precisely controlled. There is limited observational data available to model this dependence, however \cite{Georgiou2019a} found a roughly linear relationship between the IA amplitude and the radial sensitivity of their shear estimator. The difference in alignment amplitude between their smallest and largest radial weightings would correspond to $a\approx0.3$. \cite{Singh2016}, when comparing \textit{Re-Gaussianisation}, \textit{Isophotal}, and \textit{de Vaucouleurs} shapes, found $20\%$ to $40\%$ differences in alignment amplitude between various combinations of estimators, which corresponds to values of $a$ from $0.6$-$0.8$, though these estimators were not optimised to result in a maximal difference in IA contamination.

It is therefore reasonable to expect an optimised estimator could achieve somewhere in the range $0.3 \leq a \leq 0.6$, though this is complicated by factors such as the variation in the distance between isophotes for different galaxy types and sizes. Such variation could mean values of $a$ differ for each galaxy within a sample and across different redshift bins (since the former properties are known to be correlated with redshift). Further investigation into this is therefore necessary in the design of optimised estimators, to ensure any variation is not large enough to seriously dilute a MEM signal.

At present, we suggest studies looking to optimise shear estimators for the MEM aim for the lowest value of $a$ possible in the design stage, and when applying these estimators to observational data, do not rely upon the assumption that the values of $a$ seen in simulations will carry through to the observational sample. A sensible first step for a successful measurement with the MEM is to seek only to constrain the scale dependence of the IA signal, which alone could provide valuable insight into the IA of galaxies.

\begin{figure*}
    \begin{subfigure}
        \centering
        \includegraphics[width=0.46\textwidth]{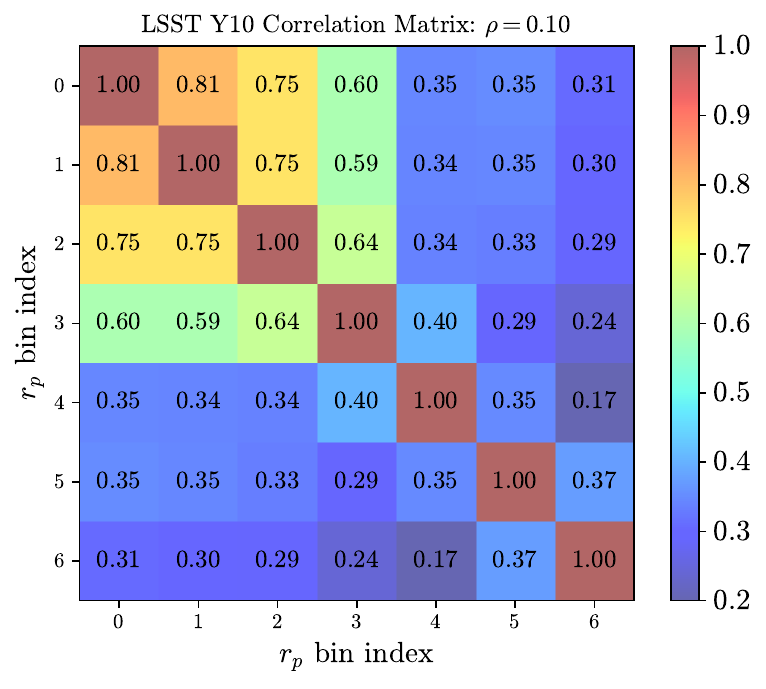}
    \end{subfigure}
    \hspace{5mm}
    \begin{subfigure}
        \centering
        \includegraphics[width=0.46\textwidth]{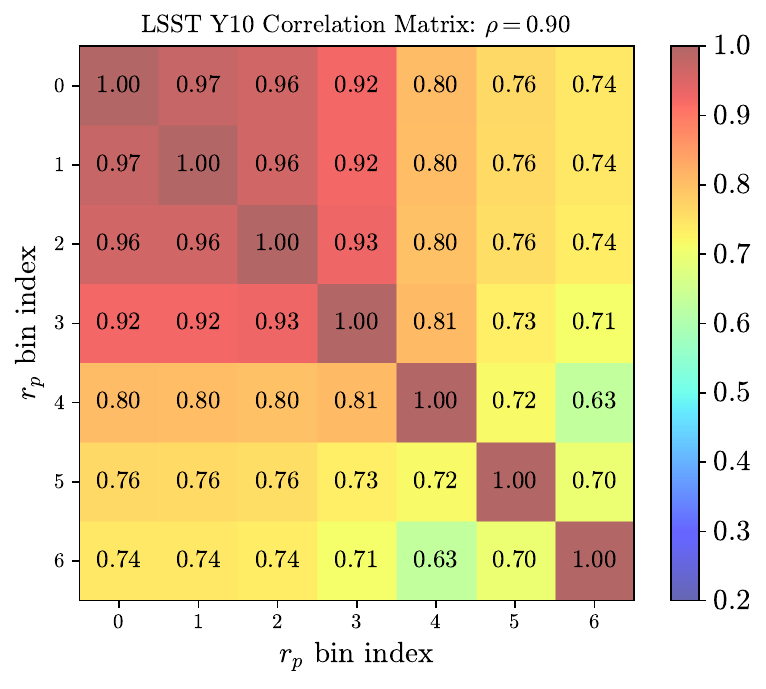}
    \end{subfigure}
    \caption{Correlation matrices for LSST Y10 at $\rho = 0.10$ (left) and $\rho = 0.90$ (right). In the right panel we see very high correlation across the entire matrix, including in the larger $r_{\rm p}$ bins. This is not the case for the lower $\rho$ value on the left, however, in the smaller $r_{\rm p}$ bins we still see significant correlation.}
    \label{fig:correlation_mats}
\end{figure*}

\begin{figure*}
    \begin{subfigure}
        \centering
        \includegraphics[width=0.48\textwidth]{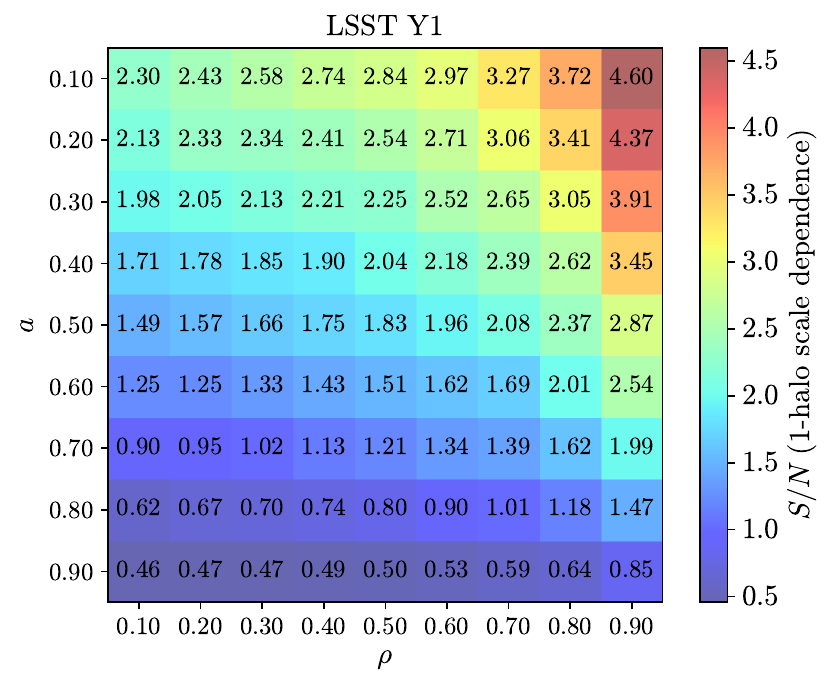}
    \end{subfigure}
    \hspace{5mm}
    \begin{subfigure}
        \centering
        \includegraphics[width=0.48\textwidth]{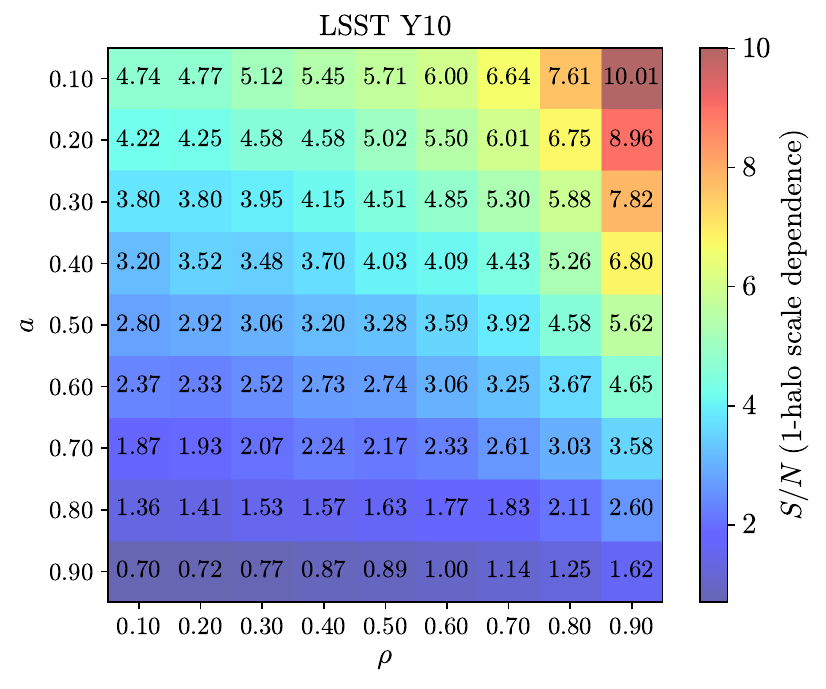}
    \end{subfigure}
    \caption{Signal-to-noise ratio for the constraint on 1-halo scale dependence with LSST Y1 (left) and Y10 (right) like data. Different combinations of $a$ and $\rho$ are shown. We define the signal as the median of the posterior distribution of $b_{\rm 1h}$ values and noise as the region containing $68\%$ of the posterior probability ($1\sigma$ uncertainty).}
    \label{fig:mcmc}
\end{figure*}

\subsubsection{Requirements for the detection of IA}
\label{sec:detect}
The simplest question of whether or not we expect a detection of IA can be answered by looking at the SNR in a single $r_{\rm p}$ bin. To claim a detection, we require (assuming all other systematics are controlled) that the SNR be greater than or equal to $1$ for a given $r_{\rm p}$ bin, else the $1\sigma$ error bars would be consistent with zero. Figure \ref{fig:snr_diag} shows the per-bin SNR as a function of $r_{\rm p}$ for Y1 and Y10, visualised for a selection of $\rho$ and $a$ pairings which are `borderline' in terms of detection.  

From Figure \ref{fig:snr_diag} we can infer that, to obtain a detection of IA in all or most $r_{\rm p}$ bins, a value of $a \leq 0.6$ is required. Higher shape noise correlation values would allow for a detection with slightly higher values up to $a \leq 0.7$, but $a$ itself appears to have a greater impact on the per-bin SNR in all cases, as shown by the wider spread of points in the bottom two panels. We therefore expect from this that, given the estimators used meet this requirement, LSST Y1 levels of multiplicative bias should still allow for a $1\sigma$ detection of IA with the MEM, with Y10 allowing for potentially a $2\sigma$ or higher detection for the same estimators. It is important to note the $a$ values considered here may be conservative. We have chosen these values to showcase clearly the boundary at which the MEM becomes unable to detect the IA signal; of course, pushing $a$ further below $0.6$ would allow for an even stronger detection. For now, we expect as long as the estimators have $a \leq 0.6$, the MEM has potential to detect IA in LSST Y1. Such a value should be achievable given the differences in IA amplitude found in~\cite{Singh2016}.

In all cases, significantly higher signal to noise is seen in the lowest projected separation bin, as a result of the different scale dependencies of the IA and lensing signals (which propagates forward into the lensing residual and thus our covariance) inside the 1-halo regime. For example,~\cite{Georgiou2019a} find the 1-halo IA signal to be represented by a power-law in $r_p$ with a scale dependence index of $b=-2$ in Galaxy and Mass Assembly survey (GAMA) and KiDS data. This value is also used by~\citetalias{Fortuna2021} in the construction of their IA halo model, and as such we have chosen to use the same value in our modelling here. On the other hand, the lensing signal scale dependence seen in similar samples used by~\cite{Viola_2015} and~\cite{Dvornik_2018} appears to follow roughly $b=-1$. This implies it is not unsurprising that at projected separations of $r_{\rm p} \approx 0.1$Mpc/h we begin to see a much stronger alignment signal, resulting in a very high SNR.

\subsubsection{Full covariance signal-to-noise}

The question of whether we expect a detection of IA is not the only one, however, particularly as the MEM is, by construction, unable to independently measure the amplitude of an IA signal (in isolation from precise external information on the value of $a$). We are thus motivated to consider the constraining power of the signal more broadly, with a key objective of the MEM being to place model independent constraints on the IA scale dependence. To move towards understanding this, we first consider the overall SNR, which can be obtained from the full covariance matrix in all $r_{\rm p}$ bins with the following equation,
\begin{equation}
    \frac{S}{N} = \sqrt{[(1-a)\bar{\gamma}_{\rm IA}]{\rm Cov}_{\rm IA}^{-1}[(1-a)\bar{\gamma}_{\rm IA}]^T}.
\end{equation}
From this definition, we can see how the covariance between $r_{\rm p}$ bins affects the total SNR across all bins. As $\rho$ impacts the covariance matrix, computational limitations mean we cannot calculate a large quantity of covariance matrices to probe $\rho$ values. Instead, we calculated the covariance matrices for nine $a$ and $\rho$ values between $0.1$ and $0.9$, resulting in $81$ combinations. We then interpolate to obtain a smooth picture of the SNR across the $a$-$\rho$ parameter space.

Figure \ref{fig:overall_snr} shows the full covariance SNR for Y1 and Y10. Across, the entire parameter space, we see very high signal to noise, even in places where we would not expect to obtain a detection of IA. There are two reasons for this. First, as discussed in Section \ref{sec:detect}, even for poor values of $a$ and $\rho$, there is still a strong signal in the lowest separation bin. Second, the covariance matrix has highly correlated off diagonal terms, particularly in lower $r_{\rm p}$ bins. This is shown in Figure \ref{fig:correlation_mats}. This high correlation implies that, while $\rho$ has a less significant effect on whether we expect a detection of IA in a given $r_p$ bin or not, its effect on the constraining power of the overall measurement with respect to parameters of interest may be significant. We will now go on to explore if this is indeed the case.

\subsubsection{1-halo scale dependence constraints}

Evidently, the overall SNR does not, by itself, provide a full picture of the forecast constraining power of the MEM, with respect to model parameters of interest, namely, scale-dependence. To explore this in greater detail, we carry out MCMC fits using the \textsc{emcee}~\footnote{\url{https://github.com/dfm/emcee}}~\citep{Foreman_Mackey_2013} package. We fit the synthetic measurement to a 4-parameter truncated power-law model, designed to qualitatively approximate our fiducial IA model, while maintaining a realistic level of model agnosticism:
\begin{multline}
    (1-a)\bar{\gamma}_{\rm IA} = a_{\rm 1h}r_{\rm p}^{b_{\rm 1h}}\left(\textrm{exp}\Bigg[-\left(\frac{r_{\rm p}}{0.3}\right)^2\Bigg]\right) \\
    + a_{\rm 2h}r_{\rm p}^{b_{\rm 2h}}\left(1-\textrm{exp}\Bigg[-\left(\frac{r_{\rm p}}{0.75}\right)^2\Bigg]\right).
    \label{eqn:trun_power_law}
\end{multline}
Note that in this case, the amplitudes which are fit ($a_{\rm1h}$ and $a_{\rm 2h}$) will be a fraction of the true amplitude, dictated by the value of $a$. Compared to equation \ref{eqn:c_ell}, we have swapped the order of the truncation terms as $\ell$ and $r_{\rm p}$ are inversely proportional. The truncation scales of $0.3$ Mpc/h and $0.75$ Mpc/h are chosen to give the best fit to the fiducial signal from maximum likelihood estimation. 

To constrain the model parameter space, we adopt a set of uniform priors, $a_{\rm 1h} \in [0,10]$, $b_{\rm 1h} \in [-10,0]$, $b_{\rm 2h} \in [-10, 0]$, and $a_{\rm 2h} \in [0,10]$. We run chains for each of the $81$ combinations of $a$ and $\rho$, initialising $32$ walkers in a small spread around the maximum likelihood estimates, and allowing the chains to run until all have achieved convergence. which we define as, $N > 50\tau$, where $N$ is the total number of iterations and $\tau$ is in the integrated auto-correlation time. We note that a stricter test of convergence should ideally be used when placing model constraints with real data, but for the purposes of probing the acceptable values of $a$ and $\rho$, this criteria is sufficient. 

Marginalising over the other 3 parameters in the model, we estimate the forecast SNR for the 1-halo scale dependence, $b_{\rm 1h}$, by taking the $50$th percentile (median) value as the best fit and the distance between the $16$th and $84$th percentiles as the $1\sigma$ ($68\%$) confidence region. We choose to take the median rather than the maximum likelihood estimate, as we found it to be more robust to variations in walker initialisation and allowed more freedom of the model within the signal uncertainties. The resulting SNR from these fits is shown in Figure \ref{fig:mcmc}. Encouragingly, we see that even for certain $a$-$\rho$ combinations where we do not expect detection, a $1\sigma$ or greater constraint on the scale dependence is still possible with high enough values of $\rho$. The importance of high $\rho$ values is further emphasised here for ensuring the tightest possible constraints, with $\rho=0.90$ resulting in twice the SNR compared to $\rho=0.10$ for the same $a$ value. Similar to what was seen in the per-bin diagonal SNR, going from Y1 to Y10, we again see an approximately factor of $2$ increase in the constraint SNR.

An interesting question also arises when we consider the relation between $a$ and $\rho$ themselves. If we could expect that $\rho$ were to increase as $a$ decreased, it would be greatly beneficial to the development of bespoke estimators. However, the inverse could make designing an estimator suitable for Y1 in particular challenging. Answering this question would require the analysis of specific shape estimators, which is beyond the scope of this work, but we highlight this as a key consideration for future research.

As a heuristic guide in interpreting the results shown in Figure \ref{fig:mcmc}, we consider the analysis of~\cite{Secco2022}, which sought to select an appropriate model of IA for the DES Year 3 cosmic shear analysis. That work found that, in the scenario where the true IA was described by the Tidal Alignment - Tidal Torquing model \citep[TATT,][]{Blazek2019} with tidal torquing amplitude $a_2=-1.36$, redshift index of this term $\eta_2=-2.5$ and source density bias parameter $b_{\rm TA}=1.0$, but the analysis incorrectly assumed the simpler NLA-$z$ model with $a_2=\eta_2=b_{\rm TA}=0$, the result was a bias in the $S_8 - \Omega_{\rm M}$ plane of considerably more than $2\sigma$. We can see that for each of these assumed truth values, an external measurement with SNR of $1$ would have excluded the incorrectly assumed value at the $1\sigma$ level, while a prior measurement with SNR of 2 would have decidedly ruled out the possibility of using the simpler presumed model. Thus, while a higher SNR is of course helpful for further pinning down model properties, a measurement of an IA model parameter of SNR even modestly >1 can be extremely powerful in ensuring robust cosmological constraints from a cosmic shear analysis, further emphasising the importance of improving our ability to measure and constrain IA in an observational context.

\section{Discussion and Conclusions}
\label{sec:conclusion}

In this work, we have carried out the first application of the Multi-Estimator Method (MEM) for measuring and / or constraining intrinsic alignment (IA) developed in~\cite{Leonard2018} (\citetalias{Leonard2018}). Using Dark Energy Survey Year 1 (DES Y1) shear estimators, we showed how the MEM could be applied, and investigated and corrected for systematic errors that may pose problems to attempts to use this method. We identified three key systematics that future applications must treat or account for in order for the technique to succeed:

\begin{itemize}
    \item Selection biases induced when making catalogue cuts to match the lensing contribution to shear between shape samples.
    \item Differences in effective weighting schemes between the two samples, altering the effective redshift distribution of the samples, and thus the measured lensing signal.
    \item Residual multiplicative biases in the lensing signal, due to calibration uncertainty, resulting in a lensing residual when cancellation is performed.
\end{itemize}

Our investigation into these systematics highlighted the potential biases that can be introduced when they are not accounted for and we therefore attempted to develop techniques to treat them. Given we expected our uncertainties to be underestimated, the final result was a null detection of IA in DES Y1, which was qualitatively consistent with the findings of~\cite{Samuroff2019}. We found the dominant source of uncertainty in this context was statistical. Having developed the tools and knowledge necessary to apply the MEM in an observational context, we went on to forecast the significance of the multiplicative bias induced lensing residual in Stage IV data.

We additionally considered the requirement on galaxy size, such that a galaxy should be well enough resolved that differences in radial weighting between two shape estimators are physically meaningful. Using a strict cut on galaxy effective size of $R \geq 3$, we constructed new samples from the original forecasts for the Legacy Survey of Space and Time  (LSST) Y1 and Y10 source samples. We used halo occupation distribution models to theoretically determine the observed quantities necessary for the MEM, and thus forecast IA and lensing residual signals.

For a fiducial IA signal in our forecasts, we used a combination of the IA halo model (\citealp{Schneider_2010}; \citetalias{Fortuna2021}) and a redshift dependent non-linear alignment (NLA-$z$) model with parameters from the DES Y3 best fits~\citep{Secco2022}. We stress again, this choice, while well motivated by observations and literature, may not necessarily be representative of the contamination in LSST, and therefore our forecasting results are guidelines for future applications of the MEM, rather than strict requirements. We plan in future work to determine how varying IA models could impact the signal obtained from the MEM via analysis of simulated galaxy images.

With a set of fiducial signals, we developed a scheme whereby uncertainty in the multiplicative bias can be accounted for as a systematic error and estimated its contribution to the covariance. We found even in the presence of multiplicative bias uncertainty $m = \pm0.013$, there was possibility of a $1\sigma$ or higher detection in LSST Y1, given the offset in alignment amplitude between the two estimators is greater than $40\%$ (represented by the MEM parameter $a \leq 0.6$). In this case, the impact of shape noise correlation between estimators (captured by the Pearson correlation coefficient, $\rho$) was limited. However, higher $\rho$ values became more important when the signal-to-noise on the signal was close to $1$, although $a$ in all cases still remained the primary factor in determining if detection was possible. For LSST Y10, the drop in multiplicative bias uncertainty to $m = \pm 0.003$ resulted in a roughly factor of $2$ increase in the SNR, and thus enabled detection at higher $a$ values. A general limit for Y10 could be taken as $a \leq 0.80$.

When addressing the ability of the MEM to constrain the 1-halo scale dependence of the IA signal, we found $\rho$ became more significant, due to high covariance in the forecasts. With sufficiently high $\rho$, constraints on the scale dependence could have a SNR greater than $1$ even if the data itself was consistent with zero. We also found for low SNR values of $a$ and $\rho$ that the maximum likelihood estimate from the chains often over-fit the data points themselves, despite the presence of large uncertainties, whereas the median values of the posterior distribution allowed for more freedom of the model within the error-bars. For this reason, we chose to use the median as our best fit value when calculating the 1-halo scale dependence constraint SNR.

As a general guideline, for high values of $a$, achieving a value of $\rho = a$ should be sufficient to obtain a reasonable constraint on the 1-halo scale dependence. This requirement relaxes for lower $a$ values, allowing for lower values of $\rho$. Given the results here, we recommend a realistic and achievable target for LSST Y1 is $a=0.60$ and $\rho=0.50$. Given the model used here, this would allow for both a detection and greater than $1\sigma$ constraint on the scale dependence. Exceeding this baseline would of course result in even more favourable performance of the MEM.

In future, we will look to identify specific shear estimators and optimise them for use with the MEM by introducing custom radial weighting. As mentioned previously, it would be interesting to investigate how values of $a$ and $\rho$ are related for different pairs of estimators, to determine the feasibility of achieving both low $a$ and high $\rho$. This is a study that is most easily carried out in tangent to shear estimator optimisation. Furthermore, we will seek to minimise multiplicative bias uncertainty in the estimators as much as possible, in the hopes of lowering it well beyond the LSST Y1 requirements. 

Promising shear estimators for optimisation include \textsc{METADETECT}~\citep{sheldon2023}, which builds upon the framework of \textsc{METACALIBRATION} to also perform galaxy detection, and \textsc{Fourier Power Shapelets}~\citep{Li_2018,Li_2020,Li_2022}, which has analytical correction for measurement bias making it computationally efficient in the context of the MEM, where a second shear estimation pipeline needs to be run on images. Finally, \textsc{Forklens}~\citep{zhang2023} is potentially also promising, due to its ability to measure shear from extremely noisy images. This may remove the need for galaxy weights entirely, therefore circumventing the requirement for a matched weighting scheme.

In conclusion, here we have built upon the work of \citetalias{Leonard2018} to develop a greater understanding of the systematics present in the MEM, and how they can potentially be treated or accounted for. We have carried out the first application of this method to observational data using DES Y1 and measured an IA signal consistent with zero, that includes the DES Y1 best fit NLA model within its uncertainties. In the context of LSST Y1 and Y10, we have placed general requirements on the key parameters relating the shear estimators used, in order  ensure future attempts at measurement are robust to contamination by residual multiplicative bias. Our work here has shown that, while challenges lie ahead, the measurement of IA in LSST Y1 is possible and strong constraints in Y10 are highly likely, especially if development of the MEM continues for LSST Y1 and beyond. To further develop the guidelines given here, work identifying and optimising shear estimators for the MEM is required and planned for future analysis. Tests on simulated galaxy images with these tailored estimators can be used to determine how the fiducial IA model affects the method, and allow specific requirements on galaxy size in the context of specific values of $a$ to be placed. With the work carried out here and proposed for the near future, the MEM has the potential to become another important tool in developing our understanding of the intrinsic alignment of galaxies.

\section*{Acknowledgements}

We thank Christos Georgiou, Joachim Harnois-D\'{e}raps, Carlos Garcia-Garcia, Joe Zuntz, Jonathan Blazek, Benjamin Joachimi, and Javier Sanchez for helpful discussions. Thanks also to Fran\c{c}ois Lanusse, Danny Dixon, Devang Liya, and Aaron Piccolo for advice on machine learning techniques, and to Chris Harrison for kindly allowing us to use his computing resources. Thanks to the Lorentz Center, Leiden, for hosting a workshop where some of the discussions were held and work carried out. CMG is supported by the Newcastle University Lady Bertha Jeffreys studentship. This work would not have been possible without the Python libraries \textsc{SciPy}~\citep{2020SciPy-NMeth}, \textsc{NumPy}~\citep{Numpy}, \textsc{PyCCL}~\citep{Chisari_2019}, \textsc{TreeCorr}~\citep{Jarvis2015}, \textsc{TJPCov}~(\url{https://github.com/LSSTDESC/TJPCov}), \textsc{scikit-learn}~\citep{scikit-learn}, and \textsc{keras}~\citep{chollet2015}. For the purpose of open access, the author has applied a Creative Commons Attribution (CC BY) licence to any Author Accepted Manuscript version arising from this submission.

Author contributions: CMG performed the majority of analysis (writing and validating code, derivations for the covariance and power spectra, writing this paper). CDL provided the conceptual framework of the analysis, guidance on the direction of the work, advice and support, some of the code used in the analysis, and editorial support for the text.

\section*{Data Availability}

The DES Y1 data used in this analysis is publicly available at \url{https://des.ncsa.illinois.edu/releases/dr1}. The other data used in the analysis will be shared if reasonably requested in correspondence with the author.



\bibliographystyle{mnras}
\bibliography{refs} 




\appendix

\section{Selection response estimation with machine learning}
\label{app:machine_learning}

\begin{figure}
    \centering
    \includegraphics[width=0.45\textwidth]{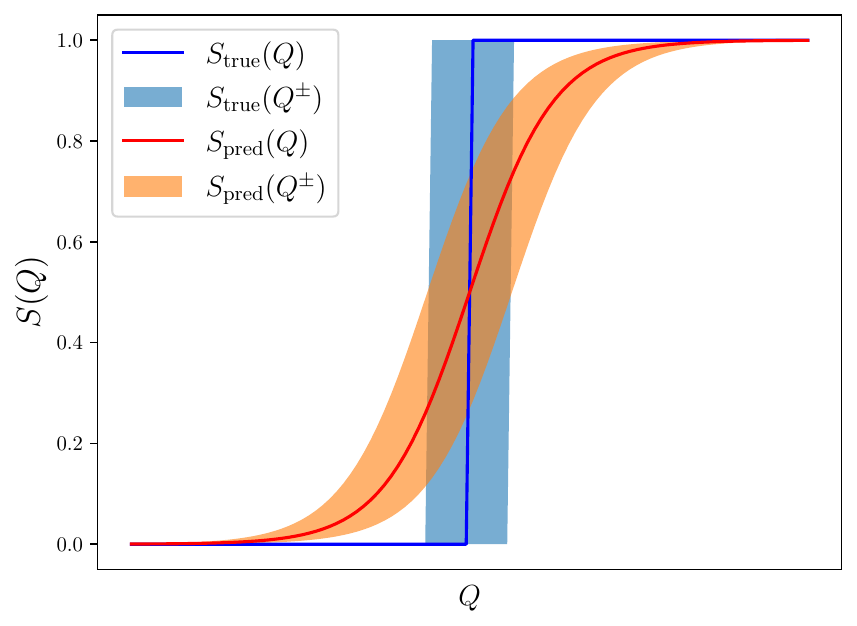}
    \caption{A hypothetical illustration of the true selection for the matched catalogue (blue), and the selection probabilities obtained from the multilayer perceptron classifier (red). The shaded regions represent how these selections might shift if they were instead made on artificially sheared image parameters. We see that a larger number of galaxies have non-zero selection probability in the classifiers predictions, thus leading to an underestimated value of the true selection response.}
    \label{fig:selection_curve}
\end{figure}

The features used by the classifier described in Section \ref{sec:response} were chosen by training and testing the classifier with various combinations of galaxy parameters and removing those which had little to no effect on classifier accuracy. To train and test the classifier, we used the full \texttt{MCAL} catalogue, with a train-test-validation split of $90/5/5$. We saw very little difference between the test and validation accuracy, indicating the classifier was not over-fit to the training data and was able to generalise well. We experimented with various architectures, but were unable to exceed a classification accuracy of $72\%$ for a threshold of $0.5$, despite achieving values above $70\%$ for a variety of model architectures, optimisers, and learning rates. For this reason, we expect that the aleatoric uncertainty in the estimation is fully controlled, and the model is achieving the maximum possible accuracy given the data available. Therefore, the imperfect classification is primarily a result of epistemic uncertainty, i.e. we lack all the information necessary to fully reconstruct the true selection function.

The final network used had a hidden layer architecture of $(8,32,128,8)$ nodes. Each layer used a rectified linear unit activation function and had an L2 regularisation term of $0.0001$ applied. We used the \textsc{Adam}~\citep{kingma2017} optimiser with a learning rate of $0.03$ and a decay rate of $0.9$ every $1000$ training steps. Each training epoch comprised of $441$ steps and training continued until the binary cross-entropy loss stopped improving by at least $1\times10^{-4}$ for $30$ consecutive training epochs, which resulted in a total of $130$ training epochs.

To understand how to interpret the value of selection response determined from the classifier's predictions, we can consider a hypothetical parameter, $Q$, which itself is a function of the galaxy parameters upon which the selection depends. In this case, the true selection function $S_{\rm true}(Q)$ can be thought of as a simple cut $Q > Q_{0}$ such that galaxies meeting this criteria have a probability of being selected equal to $1$. On the other hand, the selection function estimated by the classifier, $S_{\rm pred}(Q)$, does not include all the necessary galaxy parameters to fully reconstruct $Q$, and so a galaxy instead has some probability between $1$ and $0$ of being selected, as a result of the classifier not having access to all the parameters necessary to accurately determine $Q$. An illustration of this is shown in Figure \ref{fig:selection_curve}, where the horizontal axis should be taken to represent the range of all $Q$ values present in the full \texttt{MCAL} catalogue.

We can see from Figure \ref{fig:selection_curve} that, if we were to determine the mean ellipticity of galaxies selected from the true selection, galaxies with $Q < Q_{0}$ contribute nothing to the mean. However, in the case of the approximate selection, these galaxies would still have some weighted contribution to the mean ellipticity, dependent on the estimated probability of being selected. Therefore, the classifier provides a lower bound on the selection response, as the mean ellipticty of galaxies selected from $S_{\rm true}(Q)$ will differ more greatly from that of the full sample than the mean ellipticity of galaxies weighted by their selection probabilities from $S_{\rm pred}(Q)$.

All model architectures tested were consistent in returning negative values of selection response, and for this reason we believe the sign to be correct, with the true response most likely being some larger negative value. Models with higher accuracy (lower loss) also resulted in larger negative responses, reinforcing this belief.

\section{2-point cumulants for halo model power spectra}
\label{app:power}

\begin{figure}
    \centering
    \includegraphics[width=0.45\textwidth]{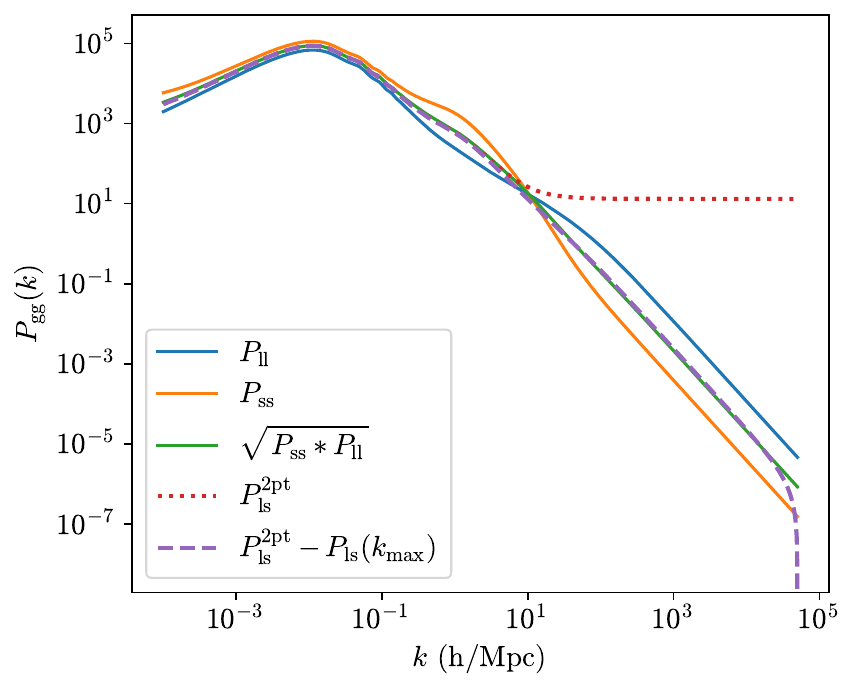}
    \caption{Comparison of galaxy-galaxy power spectra at $z=0$, calculated using the halo model with different lens and source HODs. We see the simple 2-point cumulant follows the geometric mean at low k, but begins to diverge above $k=10$. Subtracting the value at $k_{\rm max}$ from the rest of the spectrum, it once again follows the geometric mean.}
    \label{fig:power_spectra}
\end{figure}

To calculate the boost factors, we require the galaxy-galaxy power spectrum $P_{\rm gg}(\ell)$. Because we have different HODs for our source and lens samples, computing their cross-spectrum is non-trivial. Therefore, we opt to individually calculate $P_{\rm ll}$ (lens-lens clustering) and $P_{\rm ss}$ (source-source clustering), then approximate the lens-source clustering power spectrum as the geometric mean of the auto-spectra,
\begin{equation}
    P_{\rm ls}(k) = \sqrt{P_{\rm ll}(k)*P_{\rm ss}(k)}.
\end{equation}
To verify this is a suitable approximation, we plot each of the three spectra mentioned above, as well as an additional estimate, $P_{\rm ls}^{\rm 2pt}$, which is obtained by assuming the 2-point cumulant is simply the product of the means of each Fourier space profile. This plot is shown in Figure \ref{fig:power_spectra}. We find the geometric mean closely follows this 2-point term up to high $k$, at which point the spectrum becomes dominated by a constant, non-physical central-central 1-halo term, resulting from the inappropriate 2-point cumulant. Subtracting the value where this constant is most dominant (taken to be the value of $P_{\rm ls}^{\rm 2pt}$ at max $k$) from the spectrum recovers the behaviour of the geometric mean, confirming this is indeed a suitable approximation. 
For the 1-halo galaxy-intrinsic correlation, which gives the tangential shear due to IA, we derive the 1-halo power spectrum to determine if using the product of the profile means is sufficient, or if a more complex cumulant is required. The HOD for lens galaxies is given by,
\begin{equation}
    \langle v_{\rm L}(k,M) \rangle = \bar{n}_{\rm L}^{-1}\left[N_{\rm C} + N_{\rm S}u_{\rm S}(k,M)\right],
\end{equation}
where $M$ is the mass of the halo, $\bar{n}_{\rm L}$ is a normalisation over the total number of lens galaxies, and $N_{\rm C}$ and $N_{\rm S}$ are respectively the number of central and satellite galaxies for a halo of mass $M$. $u_{\rm S}(k,M)$ is a truncated Navarro-Frenk-White (NFW) profile~\citep{Navarro1996,Navarro1997}, which we assume the satellite galaxies approximately trace. From~\citetalias{Fortuna2021}, the satellite intrinsic shear profile is given by,
\begin{equation}
        \langle v_{\rm I}(k,M) \rangle = \bar{n}_{\rm s}^{-1}N_{\rm S}|\hat{\gamma}^{\rm I}(k,M)|u_{\rm S}(k,M),
\end{equation}
where I indicates the intrinsic shear, $\bar{n}_{\rm s}$ is the source satellite fraction, and $|\hat{\gamma}^{\rm I}(k,M)|$ is the part of the density weighted shear which depends on the separation of the source and lens galaxy. If we take the 2-point cumulant to be the product of the mean profiles given above, the galaxy-intrinsic power spectrum is then,
\begin{equation}
     P_{\rm gI}^{\rm 1h}(k) = (\bar{n}_{\rm s}\bar{n}_{\rm L})^{-1} \int dM \, n(M) \langle v_{\rm L}(k,M) \rangle \langle v_{\rm I}(k,M) \rangle,
\end{equation}
\begin{multline}
    P_{\rm gI}^{\rm 1h}(k) = (\bar{n}_{\rm g}\bar{n}_{\rm L})^{-1} \int dM \, n(M) \Big[N_{\rm C}N_{\rm S}u_{\rm S}(k,M)|\hat{\gamma}^{\rm I}(k,M)| \\
    + N_{\rm S}^2 u_{\rm S}^2(k,M) |\hat{\gamma}^{\rm I}(k,M)| \Big].
    \label{eqn:shear_power}
\end{multline}
From this we can infer that the product of the profile means is sufficient for the 2-point cumulant, as all terms in equation \ref{eqn:shear_power} are physically meaningful. The first term in the square brackets represents the correlation of central position with satellite shear, and the second term represents the satellite position - satellite shear correlation. We do not have a profile currently for the central shear, however, within the 1-halo regime (where this power spectrum is used in our IA model), we expect this term to be subdominant.

\section{Boost factor and F calculations}
\label{app:boost}

In order to calculate the boost factor and $F$ for the forecasting in Section \ref{sec:forecasting}, we use the theoretical expressions given in \citetalias{Leonard2018}. The boost factor is given by,
\begin{multline}
    B(r_{\rm p}) - 1 = \Bigg(\int dz_{\rm l} \frac{dN}{dz_{\rm l}} \int^{z_+(z_{\rm l})}_{z_{-}(z_{\rm l})} dz_{\rm ph} \Tilde{\omega}(z_{\rm ph},z_{\rm l}) \\
    \times \int dz_{\rm s} \frac{dN}{dz_{\rm s}} \xi_{\rm ls}(r_{\rm p},\Pi(z+{\rm s});z_{\rm l})p(z_{\rm s},z_{\rm ph}) \Bigg) \\
    \times \Bigg(\int dz_{\rm l} \frac{dN}{dz_{\rm l}} \int^{z_+(z_{\rm l})}_{z_{-}(z_{\rm l})} dz_{\rm ph} \Tilde{\omega}(z_{\rm ph},z_{\rm l}) \int dz_{\rm s} \frac{dN}{dz_{\rm s}} p(z_{\rm s},z_{\rm ph}) \Bigg)^{-1},
\end{multline}
where subscripts $z_{\rm l}$, $z_{\rm s}$, and $z_{\rm ph}$ denote lens, spectroscopic source, and photometric source redshifts, respectively. $\xi_{\rm ls}$ is the lens-source position correlation function, measured at 2D separations $r_{\rm p}$ and line of sight separations $\Pi$. $z_+$ and $z_{-}$ represent the maximum and minimum redshifts at which a source is considered to be clustered with a lens at $z_{\rm l}$. $\Tilde{\omega}$ are the galaxy weights, and $p(z_{\rm s},z_{\rm ph})$ is the probability a source at $z_{\rm s}$ is photometrically assigned a redshift $z_{\rm ph}$, and thus quantifies the photometric smear in the sample. The expression for $F$, is:
\begin{equation}
    F(r_{\rm p}) = \frac{\int dz_{\rm l} \frac{dN}{dz_{\rm l}} \int^{z_+(z_{\rm l},r_{\rm p})}_{z_{-}(z_{\rm l},r_{\rm p})} dz_{\rm ph} \Tilde{\omega}(z_{\rm ph},z_{\rm l}) \int dz_{\rm s} \frac{dN}{dz_{\rm s}}p(z_{\rm s},z_{\rm ph})}{\int dz_{\rm l} \frac{dN}{dz_{\rm l}} \int dz_{\rm ph} \Tilde{\omega}(z_{\rm ph},z_{\rm l}) \int dz_{\rm s} \frac{dN}{dz_{\rm s}}p(z_{\rm s},z_{\rm ph})}.
\end{equation}
\begin{figure}
    \centering
    \includegraphics[width=0.45\textwidth]{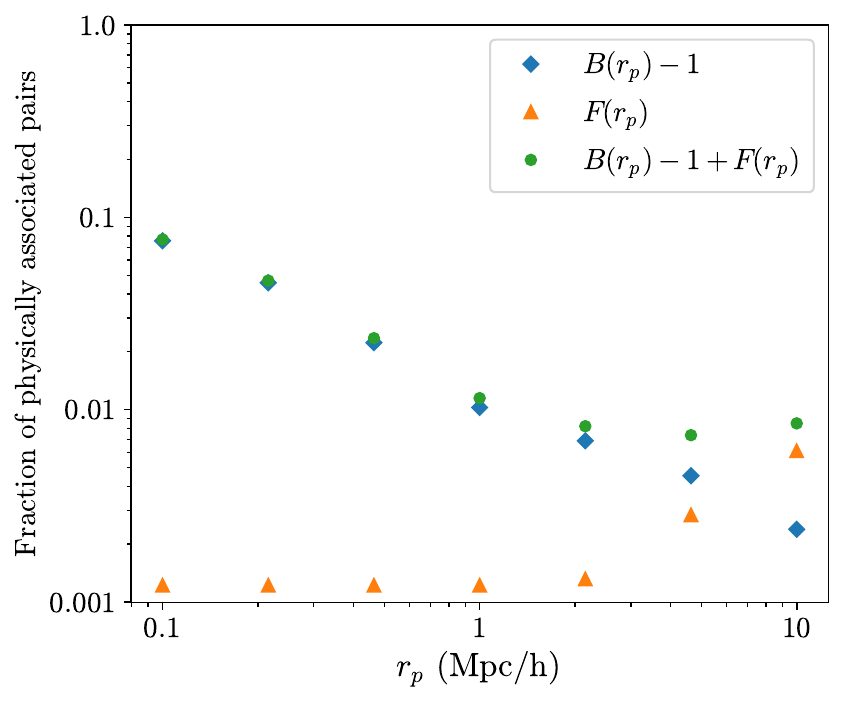}
    \caption{Boost and F values calculated using the Y1 forecasting data set detailed in \ref{sec:forecast_data}, for lenses in the range $1.0 \leq z_{\rm l} \leq 1.2$ and sources in the range $0.05 \leq z_{\rm s} \leq 2.4$. The power law for the boost follows the expected trend given in~\citep{Sheldon_2004}. We see that the inclusion of F becomes important for $r_{\rm p} > 2$Mpc/h.}
    \label{fig:boost+F}
\end{figure}
In this case, $z_+(z_{\rm l},r_{\rm p})$ represents the maximum redshift at which we expect a source to be aligned with a lens at $z_{\rm l}$, given its 2D separation from that lens is $r_{\rm p}$ (see Equation \ref{eqn:f_theta}). Figure \ref{fig:boost+F} shows the boost and F values calculated using these expressions with the LSST-like data employed in section \ref{sec:forecasting}.

\section{Derivation of covariance matrix for non-negligible multiplicative bias}
\label{app:cov}

Following the derivation in Appendix B of~\citep{Jeong_2009} and the equations given in Appendix B of \citetalias{Leonard2018}, we re-derive the expression for the statistical covariance on our measurement, in the case of non-negligible multiplicative bias. 

For two different tangential shear estimates, $\gamma_{t}$ and $\gamma_{t}^\prime$, with demonstrably subdominant post-calibration residual multiplicative bias,
the statistical covariance on our measurement can be expressed as:
\begin{multline}
    {\rm Cov}[\gamma_{\rm t}(\theta^i) - \gamma_{\rm t}^\prime(\theta^j) , \gamma_{\rm t}(\theta^i) - \gamma_{\rm t}^\prime(\theta^j)] \\ =
    {\rm Cov}[\gamma_{\rm t}(\theta^i),\gamma_{\rm t}(\theta^j)] + {\rm Cov}[\gamma_{\rm t}^\prime(\theta^i),\gamma_{\rm t}^\prime(\theta^j)] \\ - 
    {\rm Cov}[\gamma_{\rm t}(\theta^i),\gamma_{\rm t}^\prime(\theta^j)] - {\rm Cov}[\gamma_{\rm t}^\prime(\theta^i),\gamma_{\rm t}(\theta^j)],
    \label{eqn:A1}
\end{multline}
where superscripts $i$ and $j$ denote the angular separation bins in which the covariance is being calculated. Now, including the multiplicative bias residuals $m$ and $m^\prime$ in our expression and factorising on the RHS (assuming the multiplicative bias is independent of redshift and angular separation) we get:
\begin{multline}
        {\rm Cov}[(1+m)\gamma_{\rm t}^i-(1+m^\prime)\gamma_{\rm t}^{\prime, i},(1+m)\gamma_{\rm t}^{j} - (1+m^\prime)\gamma_{\rm t}^{\prime, j}]\\ 
        = (1+m)^{2}{\rm Cov}[\gamma_{\rm t}^i,\gamma_{\rm t}^j] + 
        (1+m^\prime)^{2}{\rm Cov}[\gamma_{\rm t}^{\prime, i},\gamma_{\rm t}^{\prime, j}]\\ 
        - (1+m)(1+m^\prime)\Big({\rm Cov}[\gamma_{\rm t}^i,\gamma_{\rm t}^{\prime, j}] + 
        {\rm Cov}[\gamma_{\rm t}^{\prime, i},\gamma_{\rm t}^j]\Big).
        \label{eqn:A2}
\end{multline}
An individual covariance term, ${\rm Cov}[\gamma_{\rm t}^i,\gamma_{\rm t}^{\prime, j}]$ is given by:
\begin{multline}
    {\rm Cov}[\gamma_{\rm t}^i,\gamma_{\rm t}^{\prime, j}] = \beta \Bigg[C_{\rm gM}(\ell)^{2} + C_{\rm gg}(\ell)C_{\rm MM}(\ell) \\
    + \frac{C_{\rm MM}(\ell)}{n_{\rm L}} + \rho\sigma \sigma^\prime \Big(\frac{C_{\rm gg}(\ell)}{n_{\rm s}} + \frac{1}{n_{\rm s}n_{\rm L}}\Big)\Bigg],
    \label{eqn:A3}
\end{multline}
with,
\begin{multline}
    \beta = \frac{1}{4\pi f_{\rm sky}}\int\frac{\ell d\ell}{2\pi}J_{2}(\ell \theta^i)J_{2}(\ell \theta^j).
    \label{eqn:A4}
\end{multline}
Where $f_{\rm sky}$ is the fraction of the sky covered by the survey, $C_{\rm gg}(\ell), C_{\rm gM}(\ell), C_{\rm MM}(\ell)$ are the galaxy-galaxy, galaxy-matter, and matter-matter angular power spectra respectively. $n_{\rm L}$ and $n_{\rm s}$ are lens and source galaxy surface densities respectively, $\sigma$ and $\sigma^\prime$ are the shape noise for our two estimators, and $\rho$ is the correlation in shape noise between these estimators. As such, when the estimators are the same, $\rho = 1$ and $\sigma = \sigma^\prime$.

Because, the multiplicative bias residual is assumed to be small for any contemporary shear estimators, we will neglect terms of second order in $m$ and $m^\prime$. Substituting \ref{eqn:A3} into \ref{eqn:A2} and multiplying through the brackets gives:
\begin{multline}
    {\rm Cov}[(1+m)\gamma_{\rm t}^i-(1+m^\prime)\gamma_{\rm t}^{\prime, i},(1+m)\gamma_{\rm t}^j - (1+m^\prime)\gamma_{\rm t}^{\prime, j}] \\
    = \beta \Bigg[\Big(\sigma^2+(\sigma^\prime)^2 - 2\rho\sigma\sigma^\prime\Big)\Big(\frac{C_{\rm gg}(\ell)}{n_{\rm s}}+\frac{1}{n_{\rm s} n_{\rm L}}\Big)
    \\
    + 2m\sigma^2\Big(\frac{C_{\rm gg}(\ell)}{n_{\rm s}}+\frac{1}{n_{\rm s} n_{\rm L}}\Big) + 2m^\prime(\sigma^\prime)^2\Big(\frac{C_{\rm gg}(\ell)}{n_{\rm s}}+\frac{1}{n_{\rm s} n_{\rm L}}\Big) 
    \\
    - \Big(2m\rho\sigma\sigma^\prime\Big)\Big(\frac{C_{\rm gg}(\ell)}{n_{\rm s}}+\frac{1}{n_{\rm s} n_{\rm L}}\Big) - \Big(2m^\prime\rho\sigma\sigma^\prime\Big)\Big(\frac{C_{\rm gg}(\ell)}{n_{\rm s}}+\frac{1}{n_{\rm s} n_{\rm L}}\Big)\Bigg].
\end{multline}
Simplifying this further,
\begin{multline}
    {\rm Cov}[(1+m)\gamma_{\rm t}^i-(1+m^\prime)\gamma_{\rm t}^{\prime, i},(1+m)\gamma_{\rm t}^j - (1+m^\prime)\gamma_{\rm t}^{\prime, j}] \\
    = \beta
     \Bigg[\Big(\frac{C_{\rm gg}(\ell)}{n_{\rm s}}+\frac{1}{n_{\rm s} n_{\rm L}}\Big)\Big(\sigma^2+(\sigma^\prime)^2 - 2\rho\sigma\sigma^\prime \\
    + 2m\sigma^2 + 2m^\prime(\sigma^\prime)^2 - 2m\rho\sigma\sigma^\prime 
    - 2m^\prime\rho\sigma\sigma^\prime\Big) \Bigg] \\
    = \beta \Bigg[\Big(\frac{C_{\rm gg}(\ell)}{n_{\rm s}}+\frac{1}{n_{\rm s} n_{\rm L}}\Big)\Big(\sigma^2(1+2m)+(\sigma^\prime)^2(1+2m^\prime) \\
    - 2\rho\sigma\sigma^\prime(1+m+m^\prime)\Big)\Bigg].
\end{multline}
Thus, the full expression for the covariance, separating out the shape-noise only terms which only contribute to the diagonal elements, is given by:
\begin{multline}
    {\rm Cov}[(1+m)\gamma_{\rm t}(\theta^i) - (1+m^\prime)\gamma_{\rm t}^\prime(\theta^i),(1+m)\gamma_{\rm t}(\theta^j) - (1+m^\prime)\gamma_{\rm t}^\prime(\theta^j)]\\
    = \frac{1}{4\pi f_{\rm sky}}\int\frac{\ell d\ell}{2\pi}J_{2}(\ell \theta)J_{2}(\ell \theta^\prime)\\
    \times \frac{C_{\rm gg}(\ell)}{n_{\rm s}}\Bigg[\sigma^2(1+2m)+(\sigma^\prime)^2(1+2m^\prime) - 2\rho\sigma\sigma^\prime(1+m+m^\prime)\Bigg]\\
    + \delta_{ij}\frac{\sigma^2(1+2m)+(\sigma^\prime)^2(1+2m^\prime) - 2\rho\sigma\sigma^\prime(1+m+m^\prime)}{4\pi^2f_{\rm sky}n_{\rm s}n_{\rm L}}.
\end{multline}
We can see that the inclusion of multiplicative bias residuals amounts to percent level corrections in the covariance terms. Since we include the residual multiplicative bias as a systematic uncertainty in our measurement, we therefore use the original equation of \citetalias{Leonard2018} in our forecasts.

The errors shown on the forecast data points are obtained from a covariance matrix produced using \textsc{TJPCov}. However, because \textsc{TJPCov} does not yet have the functionality to calculate the covariance for the specific quantity we are interested in here, we instead employed equation \ref{eqn:A1}. 

Four standard $\gamma_{\rm t}$ covariance matrices were obtained from \textsc{TJPCov}, with the cross-estimator covariance terms calculated using $\sqrt{\rho\sigma_1\sigma_2}$ as the shape noise, varying $\rho$. This approach does not capture any potential differences in $n_{\rm s}$ for the two estimators, however, we find it a close enough approximation for the purposes of our forecasting.


\bsp	
\label{lastpage}
\end{document}